\documentclass[twocolumn]{openjournal}
\usepackage{graphicx}   
\usepackage{amsmath}    
\usepackage{amssymb,amsfonts}    
\usepackage{orcidlink}
\usepackage{verbatim}
\usepackage{hyperref}
\usepackage{xcolor}
\usepackage{longtable}
\usepackage{booktabs}

\shorttitle{Upgrading Alpha Crucis to a seven star system}
\shortauthors{I. Waisberg \& B. Katz}

\begin{document}

\title{Upgrading Alpha Crucis to a seven star system.\\Discovery of B\MakeLowercase{b} and orbital misalignment}

\footnotetext[]{Based on observations collected at the European Southern Observatory, Chile, Program IDs 60.A-9801(U), 073.C-0337, 076.C-0503, 077.C-0547, 097.D-0156 and 0100.D-0203(D,F,H,J)}

\newcommand{\weizmann}{Department of Particle Physics and Astrophysics, Weizmann Institute of Science, Rehovot 76100, Israel}

\email{email: idelwaisberg@gmail.com}

\author{\vspace{-1.2cm}Idel Waisberg\,\orcidlink{0000-0003-0304-743X}$^{1}$ \& Boaz Katz\orcidlink{0000-0003-0584-2920}$^{2}$}

\affiliation{$^1$Independent researcher, Lambda Ophiuchi Ltda}
\affiliation{$^2$\weizmann}

\begin{abstract}
Alpha Crucis is the closest very high multiplicity massive star to the Sun. At its heart is the $4" \leftrightarrow 430 \text{ au}$ binary $\alpha^1$ (A) + $\alpha^2$ (B) Cru, which combined make up the 13th visually brightest star in the night sky. Here we make use of archival VLTI data of $\alpha$ Cru A+B in order to study its multiplicity and orbital architecture. The data spatially resolved the close (6 mas) companion in $\alpha$ Cru A (a known spectroscopic binary) and revealed that $\alpha$ Cru B is also a close (17 mas) binary, which upgrades $\alpha$ Cru to a seven star system. By combining the interferometric data with radial velocities, we solve for the full orbit of Aa+Ab and find dynamical masses $M_{Aa}=17.2\pm1.2 M_{\odot}$ and $M_{Ab}=6.8\pm0.3 M_{\odot}$. While the data on Alpha Cru B are not yet sufficient to tightly constrain all orbital parameters, we find that the orbital period is most likely 405 days (with 203 days also a possibility). The orientation of the orbital planes are sufficiently constrained to yield a mutual inclination between Aa+Ab and Ba+Bb of either $50 \pm 5^{\degr}$ or $137\pm5^{\degr}$, pointing to a dynamical formation scenario for the system. The photometric masses $M_{Ba}=12.4 M_{\odot}$ and $M_{Bb}=9.8 M_{\odot}$ together with the less massive wide component $\alpha$ Cru Ca+Cb+D yield a total mass $M\simeq52 M_{\odot}$. At larger distances, the seven-star nature of Alpha Crucis would be arguably very challenging to unveil, suggesting that the companion frequency in massive star surveys may be underestimated.
\end{abstract}


\section{Introduction}
\label{sec:introduction}

\begin{figure*}[]
\centering
\includegraphics[width=0.95\textwidth]{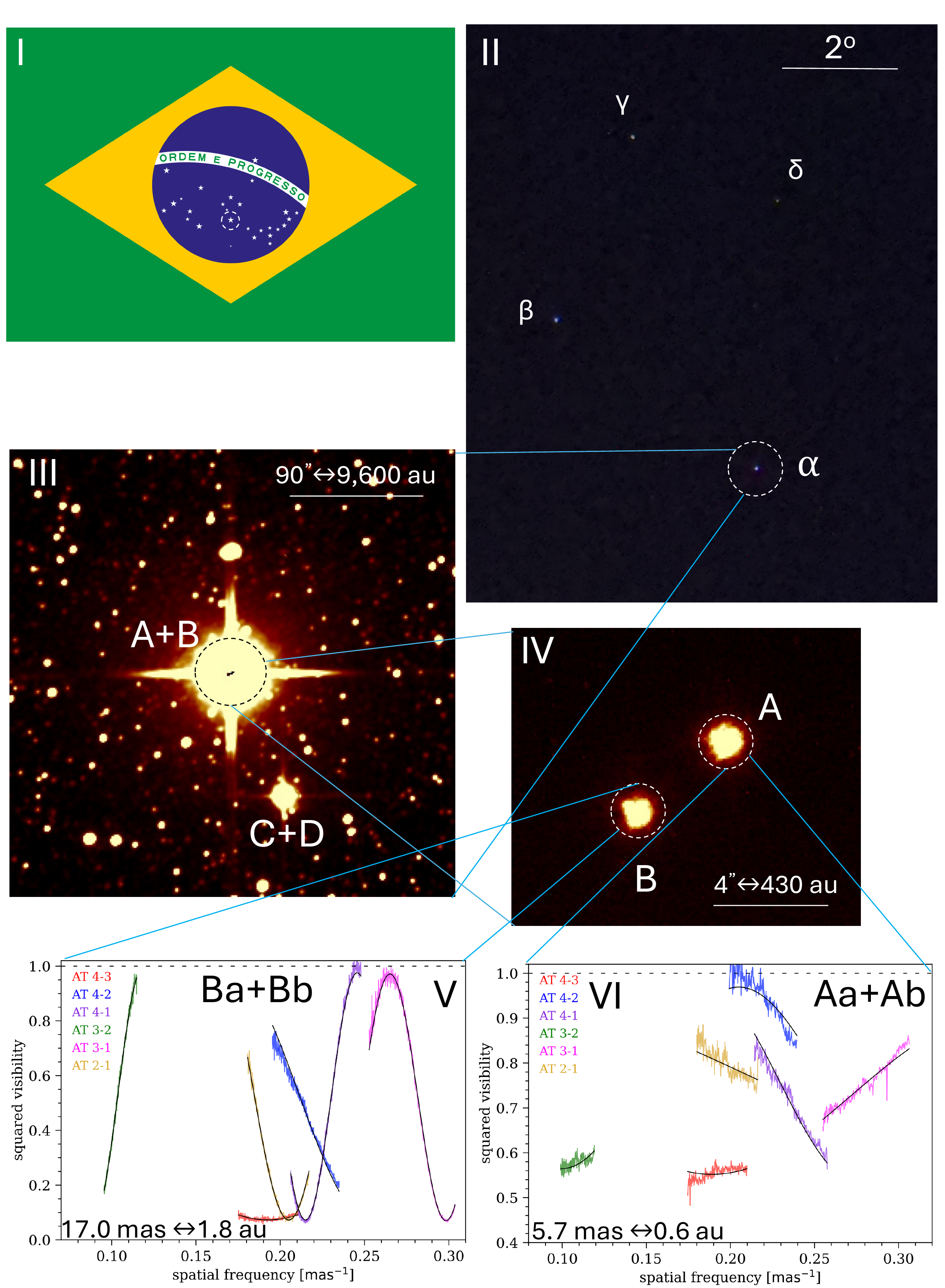}\\
\caption{\label{fig:alpha_cru_main} \textbf{I}: Alpha Crucis highlighted in the Brazilian flag. \textbf{II}: Smartphone photo of the Southern Cross as viewed from Socorro, SP, Brazil. \textbf{III}: 2MASS H band image of $\alpha$ Cru. \textbf{IV}: VLTI/GRAVITY acquisition camera H band image of $\alpha$ Cru A+B. \textbf{V/VI}: VLTI/GRAVITY interferometric data resolving $\alpha$ Cru Aa+Ab and Ba+Bb.}
\end{figure*}

Alpha Crucis ($\alpha$ Cru, HIP 60718) is the 13th visually brightest star in the night sky (V=0.80) and also the closest very high multiplicity massive star system to the Sun \citep[$\text{plx}=10.17\pm0.67 \text{ mas} \leftrightarrow d=98\pm6\text{ pc}$;][]{Hipparcos97}. As the brightest member of the Southern Cross (Figure \ref{fig:alpha_cru_main}, II), it has played a major role in many cultures of the southern hemisphere; in particular, it represents the first author's native state of São Paulo in the Brazilian flag  (Figure \ref{fig:alpha_cru_main}, I) and it also features in the logo of the European Southern Observatory. $\alpha$ Cru is also known by its IAU-approved proper name \textit{Acrux}, which now formally applies to component $\alpha$ Cru Aa. In Brazil it is also popularly known as ``Estrela de Magalhães'' after the Portuguese explorer Ferdinand Magellan. 

At its heart is the $4" \leftrightarrow 430 \text{ au}$ visual binary WDS J12266-6306, which consists of the $V_T=1.28$ B0.5IV \citep{Houk75} star Alpha1 Crucis ($\alpha^1$ Cru, $\alpha$ Cru A, HD 108248, HR 4730) and the $V_T=1.58$ B1V \citep{Houk75} star Alpha2 Crucis ($\alpha^2$ Cru, $\alpha$ Cru B, HD 108249, HR 4731). $\alpha$ Cru A is itself a single-lined spectroscopic binary \citep{Thackeray74} with period $P_A = 75.8 \text{ days}$ \citep{Thackeray80} while for $\alpha$ Cru B there are contradictory reports on whether it has \citep{Neubauer1932,Luyten1935} or does not have \citep{Thackeray74} a variable radial velocity. 

\begin{figure}[]
\centering
\includegraphics[width=0.5\textwidth]{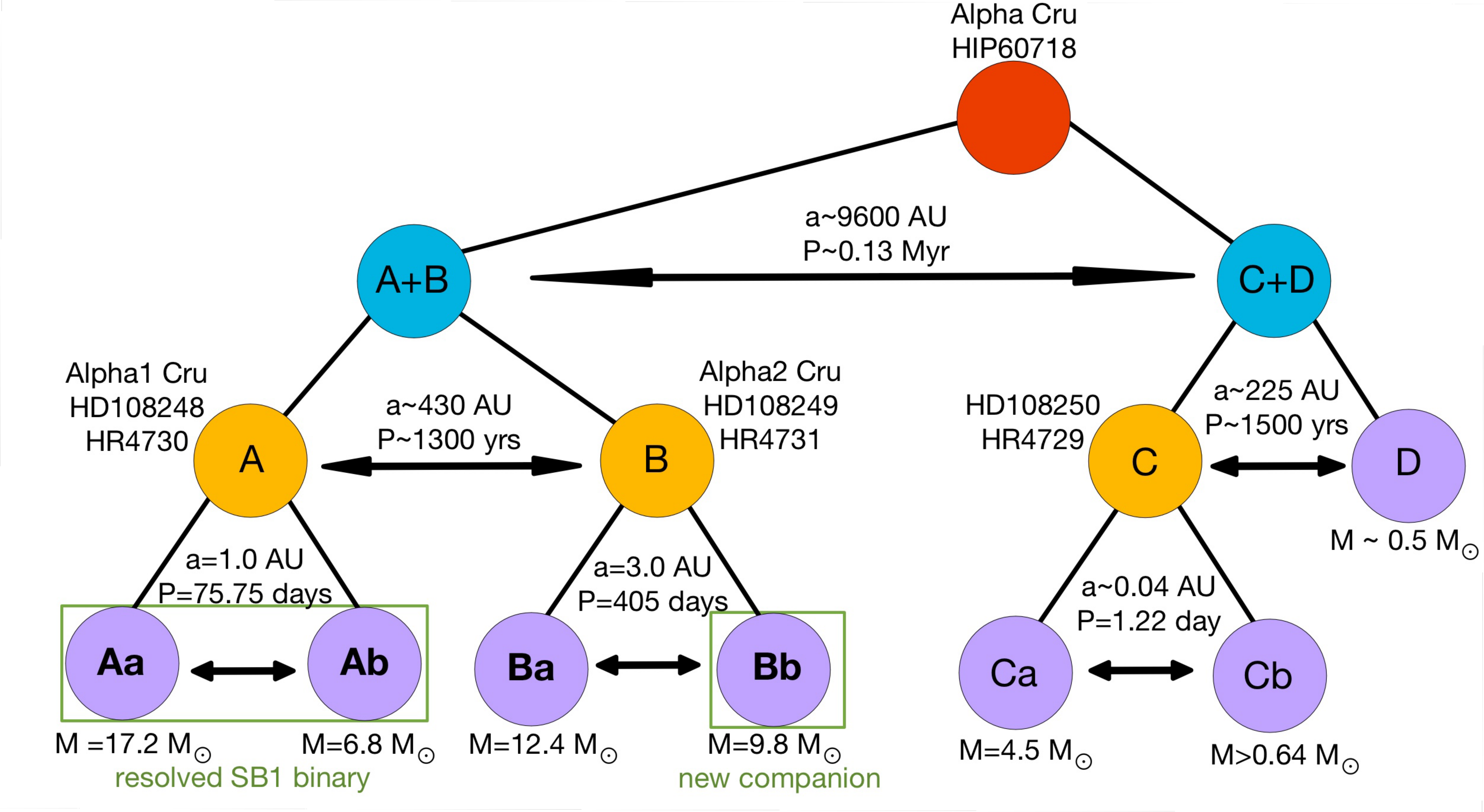}\\
\caption{\label{fig:schematic} Updated multiplicity diagram for the septuple system Alpha Crucis. The main quadruple A+B is the focus of this paper. The Ba+Bb orbit refers to the most likely period solution (see text for details).}
\end{figure}

Furthermore, $\alpha$ Cru has a wide companion (Figure \ref{fig:alpha_cru_main}, III). At a projected separation of $90" \leftrightarrow 9,600 \text{ au}$, there is a $2.1" \leftrightarrow 225 \text{ au}$ visual binary consisting of $V_T=4.8$ B4V \citep{Houk75} star $\alpha$ Cru C (HD 108250, HR 4729) and a faint $V=15.0$ M0V ($M\sim0.5 M_{\odot}$) companion $\alpha$ Cru D \citep{Tokovinin99}. $\alpha$ Cru C is itself a short period ($P_C=1.225 \text{ day}$) and circular ($e=0.024\pm0.014$) spectroscopic binary \citep{Hernandez79}. $\alpha$ Cru C was not part of the Hipparcos catalog but was solved in Gaia DR3 \citep{Gaia23}; its consistent parallax ($\text{plx}=9.37\pm0.13 \text{ mas} \leftrightarrow d=106.7\pm1.5\text{ pc}$) and similar proper motion to $\alpha$ Cru A/B leave little doubt that they are gravitationally bound. We therefore adopt the more precise Gaia DR3 distance $d=106.7\pm1.5\text{ pc}$ of $\alpha$ Cru C also for $\alpha$ Cru A+B. Figure \ref{fig:schematic} shows a schematic of the entire $\alpha$ Cru system including the results of this paper. 

$\alpha$ Cru has long been a strong candidate member of the Lower Centaurus–Crux subgroup of the Scorpius–Centaurus OB association \citep[e.g.][reported a $66\%$ membership probability.]{Rizzuto11}. More recently, the wealth of astrometric information in Gaia DR3 has allowed to study the kinematic structure of Sco-Cen in unprecedented detail; in particular, $\alpha$ Cru was found to be a member of one of the 37 different subgroups (clusters) in Sco-Cen OB2 \citep{Ratzenbock23,Ratzenbock23b}. The so-called Acrux cluster has 394 identified members at a distance of 106 pc and an age of about 11 Myr \citep[see also][for a similar conclusion]{Torosyan24}; $\alpha$ Cru is by far its brightest member, with the B3V star $\zeta$ Cru a far second. The members are spread over a large area of several square degrees; in particular, there is no other member within $5' \leftrightarrow 0.15 \text{ pc}$ of $\alpha$ Cru. 

This paper presents a deeper investigation of this iconic star based on archival VLTI interferometry and spectra. Section \ref{sec:observations} summarizes the observations. Section \ref{sec:results} presents the results, including the discovery of component Bb, a spectral analysis of $\alpha$ Cru A, the determination of orbital parameters for both subsystems and photometric masses, temperature and radii. The orbital misaligmnent and very high mulpliticy are briefly discussed in Section \ref{sec:discussion}. A conclusion can be found in Section \ref{sec:conclusion}.

\section{Observations}
\label{sec:observations}

\subsection{VLTI/GRAVITY}

We downloaded near-infrared K band VLTI/GRAVITY \citep{GRAVITY17} interferometric data on $\alpha$ Cru from the ESO archive. There are four epochs of observations, including two consecutive nights in March 2018 and another two consecutive nights in March 2019. The observations in 2018-03-03 and 2018-03-04 (PI: Domiciano) were taken in single-field mode, in which both the fringe tracker (operating at high time and low spectral resolution $R=22$) and the science fibers (at high spectral resolution $R=4000$ and integration times of 30s) are pointed at the same object. Both $\alpha$ Cru A and $\alpha$ Cru B were observed in the first epoch and only $\alpha$ Cru B on the second epoch. The observations in 2018-03-04, 2019-03-26 and 2019-03-27 are technical observatory data and were taken in dual-field mode: first $\alpha$ Cru A was used as the fringe tracker star and $\alpha$ Cru B was observed with the science fiber (at medium spectral resolution $R=500$ and integration times of 3s) and then they were swapped and observed again. All observations were made with the 1.8-meter Auxiliary Telescopes (ATs). A summary of the observations, including the mean Modified Julian Date, the seeing, AT configuration, maximum projected baseline and corresponding angular resolution, are reported in Tables \ref{table:observations_Alpha1} and \ref{table:observations_Alpha2}. 

The data were reduced with the ESO GRAVITY pipeline v.1.6.6 \citep{Lapeyrere14} and the final data include spectra, squared visibilities and closure phases (as well as differential visibilities, which are not used in this case). We always make use of the science fiber data at high or medium spectral resolution for data analysis. For the single-field data, the star HD 120913 (K2III, angular diameter 1.7 mas) was observed as interferometric calibrator. The dual-field (technical observatory) data do not include interferometric calibrators. For the epoch 2018-03-27, we used the K1III star HD 127193 (K1III, angular diameter 0.92 mas) observed about 4 hrs later in the same night as interferometric calibrator. For the epochs 2018-03-04 and 2019-03-26 there was no calibrator available. In these cases, we still make use of the data for $\alpha$ Cru B to obtain astrometric positions because the binary is well resolved; the only changes are that we fix the flux ratio to the value obtained in the other fits and include one scaling factor per baseline when fitting the squared visibilities in order to approximate the calibration (the closure phases are more robust and the calibration typically corrects for small linear slopes of the order of a few degrees, which are much smaller than the binary signatures in this case). For $\alpha$ Cru A, however, the dual data without a calibrator cannot be directly used because the binary is only partially resolved and the binary signatures are comparable to or smaller than the calibration signatures. While in principle it would be possible to calibrate $\alpha$ Cru A using $\alpha$ Cru B based on the latter's best fit binary model, we refrain from doing so because the three epochs with calibrators are already enough to constrain the orbit. Adding further two epochs (which are only one day apart from epochs already available) would have only a minor effect in measuring the orbital parameters. 

\subsection{VLTI/PIONIER}

We downloaded VLTI/PIONIER \citep{LeBouquin11} data on $\alpha$ Cru from the ESO archive (PI: Domiciano). There are two epochs for $\alpha$ Cru A (2018-02-03 and 2018-02-04) and one epoch (2018-02-06) for $\alpha$ Cru B\footnote{Formally, the target in the 2018-02-03 epoch is labeled as $\alpha$ Cru B in the file. However, we found the data are completely inconsistent with its flux ratio as measured from the other files. Instead, we found that the data are consistent with it being $\alpha$ Cru A, including the flux ratio and the position within the orbital fit.}. All observations were obtained with the 1.8-meter Auxiliary Telescopes (ATs). VLTI/PIONIER works in the near-infrared H band at low spectral resolution (in this case, with six spectral channels over the H band). The individual exposure time is 0.75 ms and each observation block consists of five files with 51,200 exposures each. The star HD 116244 (K0III, angular diameter 1.5 mas) was observed as interferometric calibrator. A summary of the observations can be found in Tables \ref{table:observations_Alpha1} and \ref{table:observations_Alpha2}.  

The data were reduced using the default settings in the PIONIER data reduction software \texttt{pndrs} v3.94 \citep{LeBouquin11}. The reduced product consists of squared visibilities for six baselines and closure phases for four triangles. For the epoch 2018-02-03, the pipeline only returns data for three baselines and one triangle because one telescope was inoperative; the available data is still enough to extract the binary parameters. 

\subsection{Historical radial velocities}

Radial velocities for $\alpha$ Cru A were taken from the tables provided in \cite{Thackeray74} (2 epochs in 1907, 3 epochs in 1914, 10 epochs in 1924, 3 epochs in 1925, 1 epoch in 1966, 1 epoch in 1969, 3 epochs in 1970 and 2 epochs in 1973) and \cite{Thackeray80} (6 epochs in 1978).

While there are 13 RVs from 1924-25 for $\alpha$ Cru B listed in \cite{Neubauer1932}, we concluded that they are not reliable. This is because the RVs for $\alpha$ Cru A listed in this same reference were very significantly revised in \cite{Thackeray74}, who had access to the same plates. The revisions were often larger than the RV values themselves and were not systematic in any way. 

\cite{Thackeray74} also lists RVs for 4 epochs from 1969-70 of $\alpha$ Cru B. These have quite large intra-epoch scatter but show no obvious change, leading the authors to conclude that $\alpha$ Cru B was a single star. However, as we will see below the interferometric data revealed $\alpha$ Cru B to be a binary with relatively low contrast, so that both components should contribute to the spectrum and the RV measured assuming a single component is highly biased (it would be constant in the case of equal components with blended lines). Therefore, there is currently no usable radial velocity data for $\alpha$ Cru B. 

\subsection{Spectra}

We downloaded ``ready-to-use'' reduced spectra of Alpha Crucis from the ESO archive. The spectra include two epochs of FEROS \citep{Kaufer1999}, two epochs of VLT/UVES \citep{Dekker2000} and one epoch of HARPS \citep{Mayor2003}. Table \ref{table:spectra} provide details of the spectral observations including dates, wavelength coverage, spectral resolution and exposure times. 

We used telluric absorption lines to confirm that the FEROS and HARPS spectra were already corrected to the heliocentric frame, while for the the UVES spectra this additional correction was needed. The spectra were normalized by fitting a cubic spline through carefully selected continuum regions. All the wavelengths are in air. 

The UVES and HARPS spectra have Alpha Cru A labeled as the target while the FEROS spectrum has Alpha Cru B. However, when studying the FEROS spectrum we noticed that it was essentially identical to the UVES and HARPS spectra of Alpha Cru A: it had the same line transitions with the same strengths and the exact same projected rotational velocity $v \sin i$. We therefore became very suspicious that the actual target observed was Alpha Cru A, which would not be very surprising given that their on-sky separation of 4" would require a dedicated observing plan to pick the correct (and in this case fainter) component. Rather definitive evidence came from the radial velocity, which (as will be shown below) is perfectly consistent with the radial velocity curve of Alpha Cru A. Therefore, we were left with no archival spectrum of Alpha Cru B.

\section{Results}
\label{sec:results}

\subsection{Discovery of Bb and separation vectors}

The VLTI/GRAVITY acquisition camera image (H band) of Alpha Cru A+B is shown in Figure \ref{fig:alpha_cru_main}, IV. The interferometric data for Alpha Crucis A  revealed a close (projected separations 5-7 mas) and relatively faint companion with an H/K band flux ratio of about $13\%$ that could be readily identified as the spectroscopic companion. Meanwhile, the interferometric data for Alpha Crucis B revealed that it is also a close (projected separations 10-17 mas) binary, with a companion that has a H/K band flux ratio of about 56\%. 

The inteferometric data was fit with a binary model \citep[detailed in][]{Waisberg23} consisting of the H/K band flux ratio of the companion and its projected separation relative to the primary in the East and North directions $(\Delta \alpha_*, \Delta \delta)$. The angular diameters of the stars were fixed to $\theta_{Aa} = 0.59 \text{ mas}$, $\theta_{Ab} = 0.28 \text{ mas}$, $\theta_{Ba} = 0.47 \text{ mas}$ and $\theta_{Bb} = 0.38 \text{ mas}$ based on the photometric radii calculated in \ref{subsec:isochrone} and have a negligible effect on the binary fits because they are well below the interferometric resolutions. The fit results for each interferometric epoch can be found in Tables \ref{table:observations_Alpha1} and \ref{table:observations_Alpha2}. 

Figure \ref{fig:alpha_cru_main} (panels VI and V) show the squared visibilities (colored) and best fit binary model (black) for the VLTI/GRAVITY data on Alpha Cru A (epoch 2018-03-03) and B (epoch 2018-03-04). Figures for all interferometric data and binary fits (including squared visibilities and closure phases) can be found in Appendix \ref{app:vlti_observations}. 

Formal interferometric fit errors are often of the order of tens of $\mu$as but are expected to be somewhat underestimated due to systematic errors and correlations between spectral channels. We therefore adopt a conservative minimum error of 0.1 mas in order to not have the orbital fit dominated by epochs with possibly underestimated errors. 

\subsection{Spectral parameters of Alpha Cru A}

\begin{figure*}[]
\centering
\includegraphics[width=\textwidth]{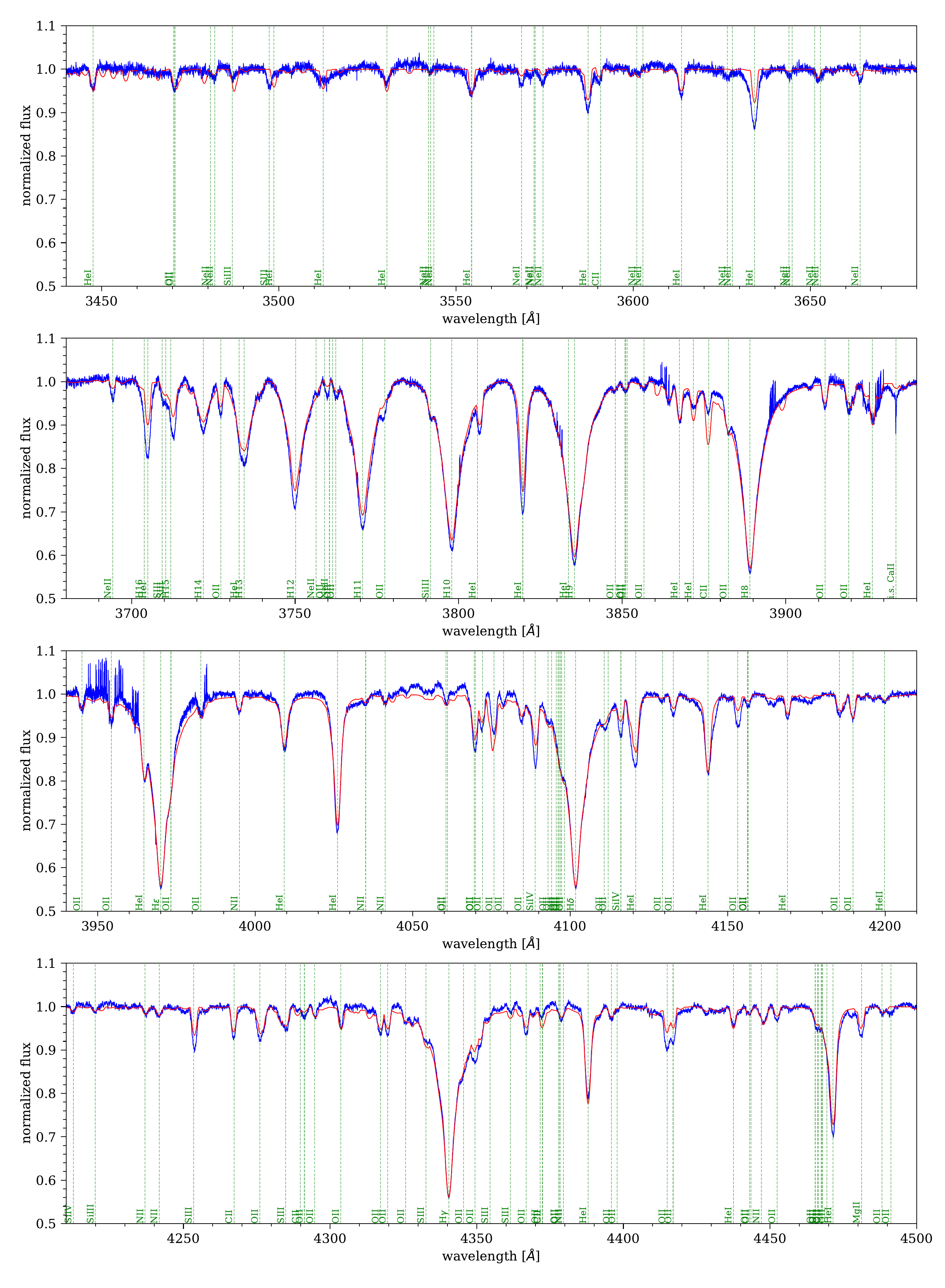}\\
\caption{\label{fig:plot_spectrum_1} Spectral atlas of Alpha Crucis A. The red line shows a  BSTAR2006 model with $T_{\mathrm{eff}}=29 \text{ kK}$, $\log g=4$ and $v \sin i = 84 \text{ km}\text{ s}^{-1}$.}
\end{figure*}

\begin{figure*}[]
\centering
\includegraphics[width=\textwidth]{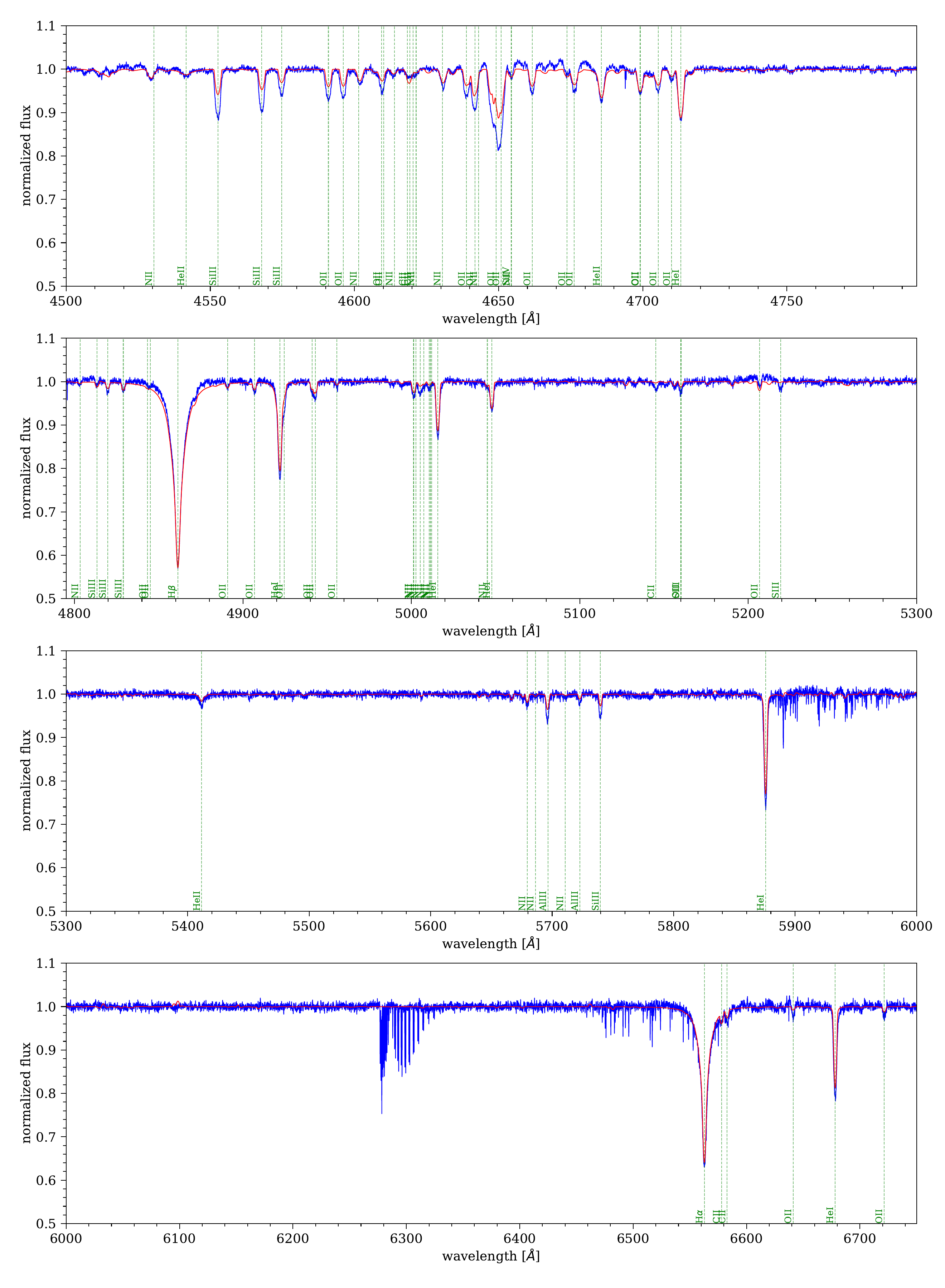}\\
\caption{\label{fig:plot_spectrum_2} Continuation of Figure \ref{fig:plot_spectrum_1}. Starting at about 5900{\AA} there are strong telluric features.}
\end{figure*}

A plot of the UVES ($\lambda<3985${\AA}) and FEROS ($\lambda>3985${\AA}) spectra (both corrected to zero velocity) is shown in Figures \ref{fig:plot_spectrum_1} and \ref{fig:plot_spectrum_2}. The strongest lines are that of He I and H I but there are also many other transitions including He II, C II, N II, O II, Ne II, Mg II, Al III, Si III, Si IV and S III. A list of the about 250 line transitions identified can be found in Table \ref{table:line_list}. There is no sign of the companion Ab, which is expected given its relatively low flux contribution and the relatively broad lines of the primary. 

The projected rotational velocity $v \sin i = 84 \text{ km}\text{ s}^{-1}$ was calculated by finding the first zero crossing of the Fourier transform of the Si III 4552.62, 4567.82 and 4574.76{\AA} lines. The effective temperature $T_{\mathrm{eff}} = 29 \text{ kK}$ was estimated by comparing the Equivalent Width ratio of the He II 4685.68{\AA} and He I 4921.93{\AA} lines with those in the publicly available NLTE, line-blanketed model atmosphere grid BSTAR2006 \citep{Lanz2007}. The grid covers the temperature range 15-30 kK in steps of 1kK and $\log g$ 1.75-4.75 in steps of 0.25 and has a solar abundance and a microturbulent velocity $\xi_t = 2 \text{ km} \text{ s}^{-1}$. The $T=29 \text{ kK}$, $\log g= 4.0$ model (after appropriate rotational and instrumental broadening and radial velocity shift) is shown on top of the data in Figures \ref{fig:plot_spectrum_1} and \ref{fig:plot_spectrum_2}. Some of the metal lines such as Si are somewhat weaker in the model compared to the data, suggesting supersolar abundances. A more detailed spectral analysis including a fit for the microturbulent velocity and individual element abundances is beyond the scope of this paper. 

Radial velocities and their errors for the primary Aa were calculated by cross-correlating the data and the template spectrum over different wavelength regions rich in absorption lines and are reported in Table \ref{table:spectra}. 

\subsection{Orbital parameters}

\subsubsection{$(P,e,T_p)$ grids}
\label{subsec:grids}

The orbital parameter space for Keplerian orbits is notoriously non-convex and requires grid approaches to find the global minimum. When dealing with astrometric data alone, the most efficient way to do this is to created three-dimensional grids in orbital period ($P$), eccentricity ($e$) and time of periastron ($T_p$) and at each point in the grid quickly compute the optimal Thiele-Innes elements $A,B,F,G$ (which map to the remaining orbital parameters, namely semi-major axis $a$, orbital inclination $i$, argument of pericenter $\omega$ and longitude of the ascending node $\Omega$) by linear regression according to

\begin{align}
\Delta \alpha_* = B X + G Y \\
\Delta \delta = A X + F Y \\
X = \cos E - e \\ 
Y = (1-e^2)^{1/2} \sin E 
\end{align}

\noindent where $E$ is the eccentric anomaly computed from Kepler's Equation. 

\subsubsection{Alpha Cru A}

In this case, we also have the radial velocities listed in \cite{Thackeray74,Thackeray80} and our own four new radial velocity measurements (Table \ref{table:spectra}). 

\begin{figure*}[]
\centering
\includegraphics[width=0.8\textwidth]{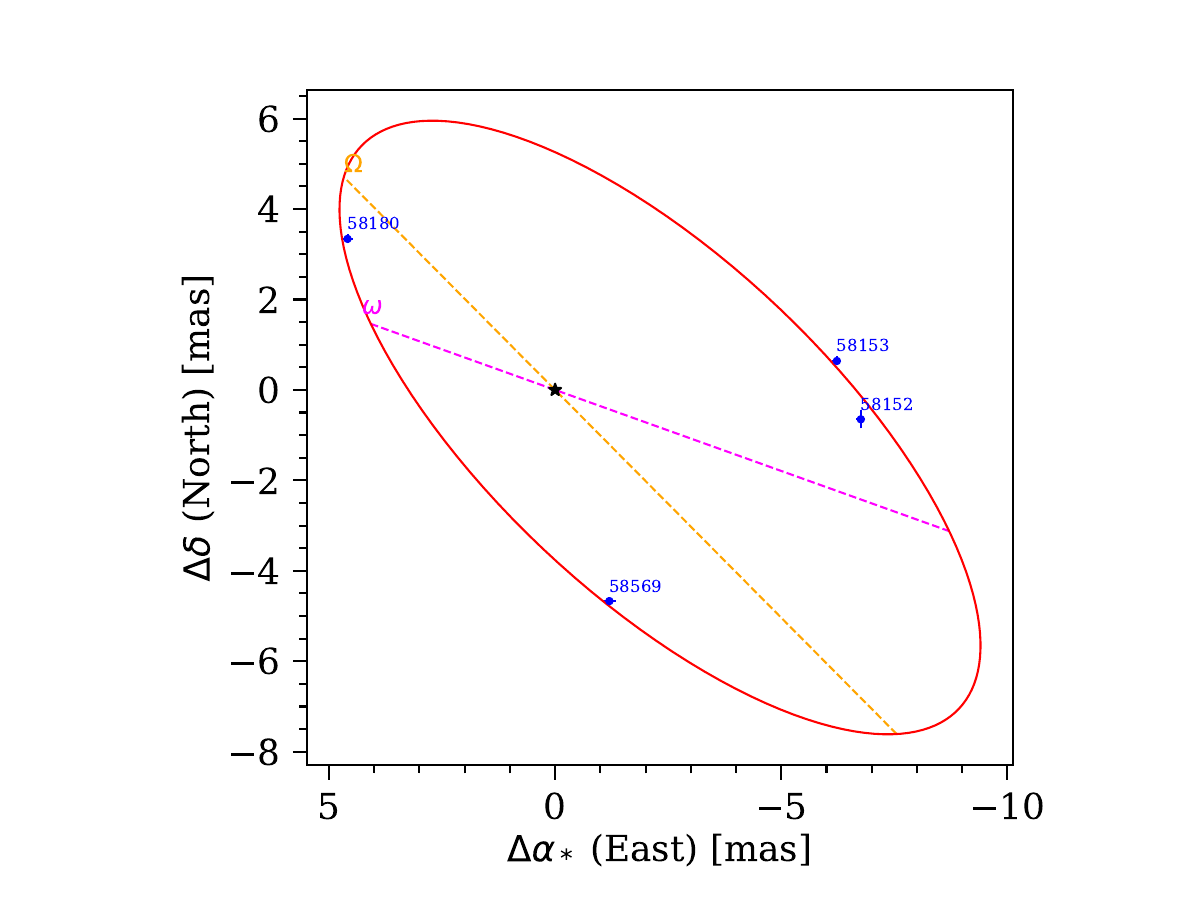}\\
\includegraphics[width=0.8\textwidth]{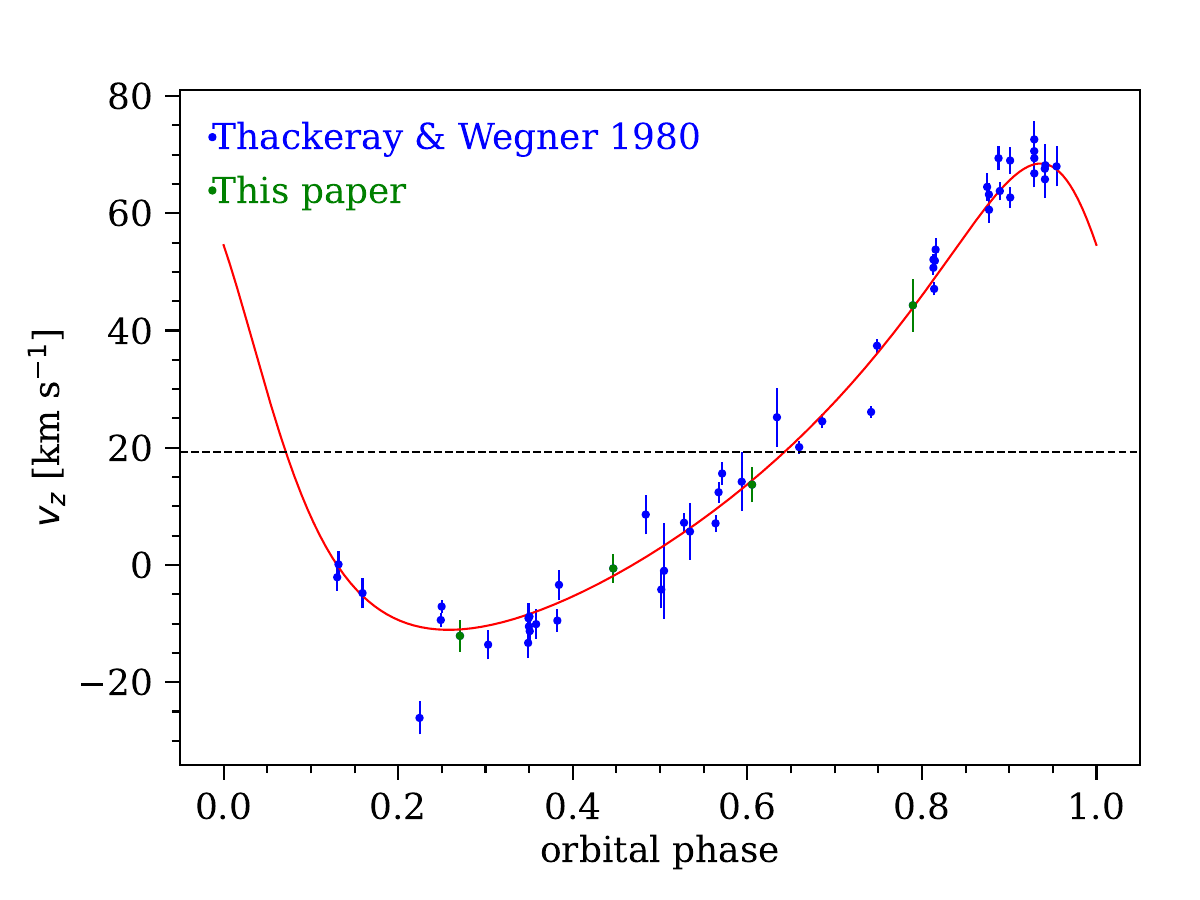}\\
\caption{\label{fig:orbits_alpha1} Data and best joint fit for the orbit of Alpha Cru Aa+Ab. \textbf{Top}: VLTI/GRAVITY and VLTI/PIONIER astrometric data (blue, labeled by MJD) and best fit orbital solution (red). The black star marks the position of the fixed primary. The dashed magenta line shows the line of apsides and the dashed orange line shows the line of nodes. \textbf{Bottom}: Radial velocity data (blue, green) and best fit orbit (red) for the primary Aa. The dashed black line shows the systemic velocity.}
\end{figure*}

In theory, a truly joint fit breaks the linearity since $v_z$ depends non-linearly on $A,B,F,G$. We adopt a hybrid approach in which we first find the optimum $A,B,F,G$ using the astrometric data alone and from them compute the corresponding $a,i,\Omega,\omega$. The radial velocity of the primary is then given by 

\begin{align} 
v_{z,a} = \gamma + K_1 (e \cos (\omega + \pi) + \cos (\omega + \pi + f)) \\
K_a = \frac{2 \pi}{P} \frac{q}{1+q} \frac{a \sin i}{(1-e^2)^{1/2}} \label{eq:rv}
\end{align}

\noindent where $q=\frac{M_b}{M_a}$ is the mass ratio, $\gamma$ is the systemic velocity and $\pi$ is added to $\omega$ because the astrometric solution always refers to the position of the secondary star relative to the primary. In practice, since the astrometric-only solution is degenerate between $(\omega,\Omega)$ and $(\omega+\pi,\Omega+\pi)$, we need to test both possible values of $\omega$ anyway. We perform a linear fit for $K_a$ and $\gamma$ (for both possible values of $\omega$) to the RV data and add the corresponding minimum $\chi^2$ to the one from the astrometric solution. This procedure yields a clear global minimum, which is then used as the starting point for a non-linear least squares fit with all parameters free to vary. In practice, this procedure is not strictly necessary in this case because the period is already well constrained from the previous RV solution. The fact that our resulting best fit period is consistent with it is nonetheless reassuring. Also reassuring is the fact that the historical RV data and our new RV data are fully consistent. We note that there was only one RV point (JD=2420604.710) listed in \cite{Thackeray74} that was an extreme outlier and it was removed from the fit. The astrometric and the radial velocity data are plotted in Figure \ref{fig:orbits_alpha1} together with the best fit solution. 

To find the orbital parameter distributions, we generated 20,000 resamples of the data according to the errors and refit the orbit. The resulting distributions of the best fit parameters are shown in Figure \ref{fig:corner_alpha1}. Table \ref{table:orbital_fits} reports the results (defined as the median of the distribution) and their uncertainties (defined as the 2.3\% and 97.7\% percentiles). The resulting dynamical mass is calculated from Kepler's Third Law and the mass ratio $q$ is calculated from Eq. \ref{eq:rv}. 

\begin{table}
\centering
\caption{\label{table:orbital_fits} Best fit orbital parameters for Alpha Crucis Aa+Ab and range of acceptable solutions for Ba+Bb for the two possible periods.}
\begin{tabular}{ccc}
\hline \hline

& Aa+Ab & Ba+Bb \\ [0.3cm]

\shortstack{$a$\\(mas)} & \shortstack{$9.48\pm0.30$\\($1.01\pm0.04$ au)} & \shortstack{$P_1: 28.2_{-0.8}^{+0.8}$\\(3.0 au)\\$P_2: 17.8_{-0.5}^{+0.5}$\\(1.90 au)}  \\ [0.3cm]

$e$ & $0.369\pm0.015$ & \shortstack{$P_1: 0.36_{-0.07}^{+0.45}$\\$P_2: 0.64_{-0.10}^{+0.17}$}  \\ [0.3cm]

\shortstack{$i$\\(deg)} & $66.2\pm1.4$ & \shortstack{$P_1: 94.4_{-0.3}^{+2.6}$\\$P_2: 96.8_{-0.9}^{+2.1}$} \\ [0.3cm]

\shortstack{$\Omega$\\(deg)} & $225.0\pm2.0$ & \shortstack{$P_1: 86.0_{-6.1}^{+3.4}$$^*$\\$P_2: 85.2_{-3.7}^{+1.7}$$^*$} \\ [0.3cm]

\shortstack{$\omega$\\(deg)} & $229.5\pm2.2$ & \shortstack{$P_1: 105.2_{-55.3}^{+32.1}$$^*$\\$P_2: 130.9_{-7.6}^{+4.4}$$^*$} \\ [0.3cm]

\shortstack{$P$\\(days)} & $75.7469\pm0.0017$ & \shortstack{$P_1 = 405.4_{-0.3}^{+0.4}$\\$P_2 =  202.9_{-0.0}^{+0.0}$} \\ [0.3cm]

\shortstack{$T_p$\\(JD)} & $2458183.1 \pm 0.4$ & \shortstack{$P_1: 2458099_{-145}^{+258}$\\$P_2: 2458091_{-17.0}^{+22.0}$} \\ [0.3cm]

\shortstack{$K_a$\\(km$\text{ s}^{-1}$)} & $40.6\pm0.9$ & - \\ [0.3cm]

\shortstack{$\gamma$\\(km$\text{ s}^{-1}$)} & $19.2\pm0.8$ & - \\ [0.3cm]

\hline \\ [0.1cm]

\shortstack{$r_p$\\(au)} & $0.64\pm0.03$ & - \\ [0.3cm]

\shortstack{$M_{\mathrm{dyn}}$\\($M_{\odot}$)} & $24.1\pm3.1$ & - \\ [0.3cm]

\hline \\ [0.1cm]

\shortstack{$q=\frac{M_{Ab}}{M_{Aa}}$} & $0.40 \pm 0.02$ & - \\ [0.3cm]

&\shortstack{$M_{Aa}=17.2\pm1.2M_{\odot}$\\$M_{Ab}=6.8\pm0.3 M_{\odot}$} & - \\ [0.3cm]

\hline
\end{tabular}
\tablenotetext{0}{Notes:}
\tablenotetext{0}{The uncertainties correspond to the 2.3\% and 97.7\% percentiles.}
\tablenotetext{0}{For the physical semi-major axis and dynamical mass, the adopted distance is $d=106.7 \pm 1.5 \text{ pc}$ ($1 \sigma$).}
\tablenotetext{0}{*: Degenerate solutions with $\Omega \rightarrow \Omega + 180^{\degr}$ and $\omega \rightarrow \omega + 180^{\degr}$ also possible.}
\end{table}

\subsubsection{Alpha Cru B}

\begin{figure*}[]
\centering
\includegraphics[width=\textwidth]{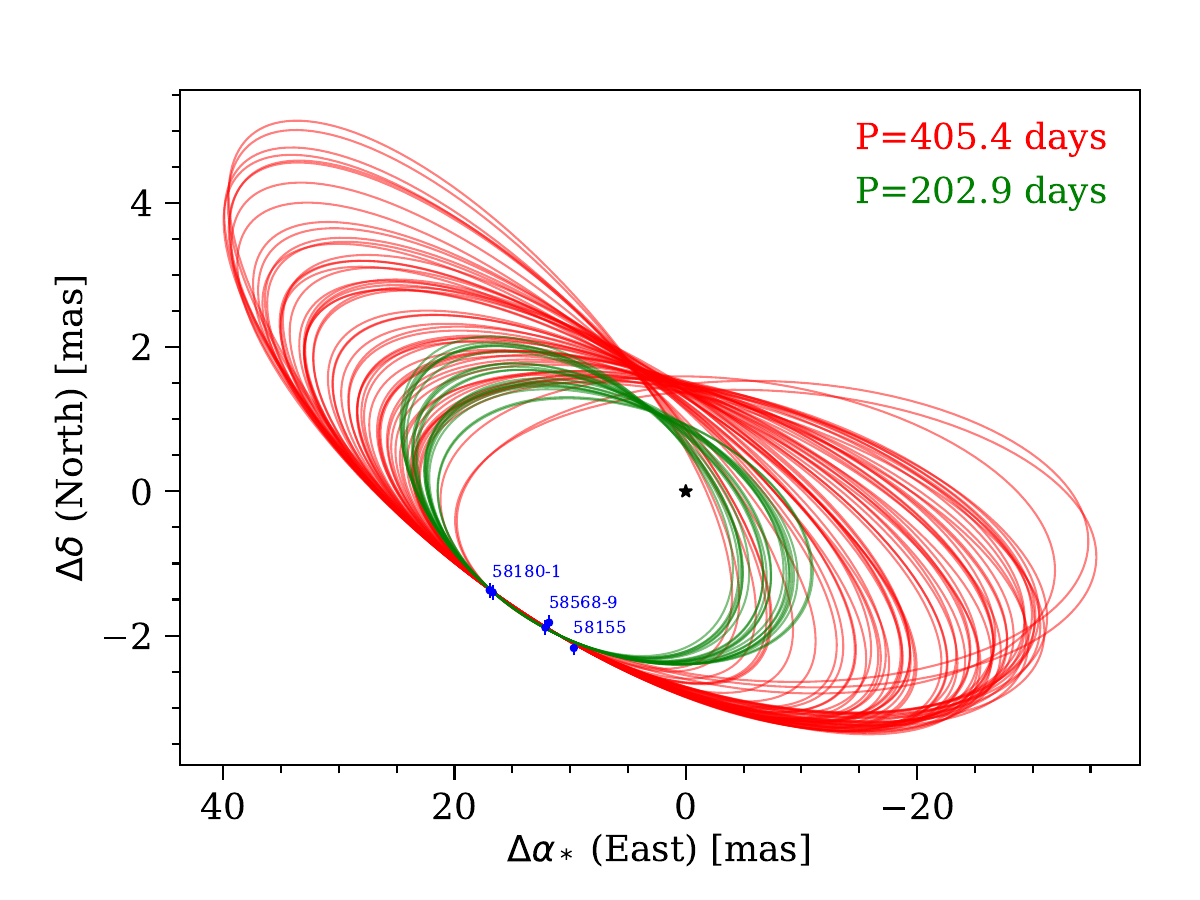}\\
\caption{\label{fig:orbits_alpha2} VLTI/PIONIER and VLTI/GRAVITY astrometric data (blue, labeled by MJD) for Alpha Crucis Ba+Bb and examples of acceptable orbital solutions for the two possible orbital periods (red and green). The dynamical mass of the binary was fixed to $20.2 M_{\odot} \leq M_{\mathrm{dyn}} \leq 24.2 M_{\odot}$. The black star marks the fixed position of the primary.}
\end{figure*}

As can be seen in Figure \ref{fig:orbits_alpha2} only a relatively small arch of the Ba+Bb orbit has been covered by the interferometric observations. While this does hinder a precise determination of all the orbital parameters, we can put meaningful constraints on them with the available data.

First, we note after the first observation (MJD=58155), the 
secondary moved about 7.0 mas in 25 days (MJD=58180-1), and then returned just past its initial position by 2.2 mas 387 days later (MJD=58568-9). Therefore, the orbital period must be of the order $387+25-\frac{2.2}{7.0}\times25 = 404 \text{ days}$ or its higher harmonics. Using the photometric mass of Ba+Bb $M_B\simeq22.2 M_{\odot}$ calculated in Section \ref{subsec:isochrone} below and Kepler's Third Law, we have the following possibilities for the period and semi-major axis 

\begin{align}
P_1\simeq404 \text{ days} ; a_1 \simeq 3.0 \text{ au} \leftrightarrow 28 \text{ mas} \\ 
P_2\simeq202 \text{ days} ; a_2 \simeq 1.9 \text{ au} \leftrightarrow 17.8 \text{ mas} \\ 
P_3\simeq135 \text{ days} ; a_3 \simeq 1.4 \text{ au} \leftrightarrow 13.5 \text{ mas} \\ 
P_4\simeq101 \text{ days} ; a_4 \simeq 1.2 \text{ au} \leftrightarrow 11.2 \text{ mas} \\ 
P_5\simeq 81 \text{ days} ; a_5 \simeq 1.0 \text{ au} \leftrightarrow 9.6 \text{ mas}
\end{align}

Given the maximum measured projected separation $\rho\sim17\text{ mas}$, shorter periods are not allowed because their semi-major axis would be too small even for a face on orbit with $e\approx1$. In practice, periods shorter than $P_3$ can also be readily excluded because in 25 days only a short portion of the orbit must be covered.

In order to find the allowed orbital parameters, we ran grids in $(P,e,T_p)$ and at each point in the grid found the corresponding optimal A,B,F,G constants as described in \ref{subsec:grids}. The remaining orbital parameters are calculated from A,B,F,G. The acceptable solutions must satisfy the following conditions: 

\begin{enumerate}
\item the resulting dynamical mass must be within 20.2 and 24.2 $M_{\odot}$ based on the photometric mass of 22.2 $M_{\odot}$. 

\item the resulting $\chi^2$ (sum of weigthed squared residuals) must be smaller than 13.3, corresponding to the 99\% percentile for a $\chi^2$ distribution with 4 degrees of freedom, that is 10 data points minus 6 parameters (the dynamical mass constraint essentially fixes $a$ for a given $P$)
\end{enumerate}

We ran three grids within $\pm 10$ days of the possible periods $P_1$, $P_2$ and $P_3$ in steps of 0.1 day, $0 \leq e \leq 0.95$ in steps of 0.01 and $-\frac{P}{2} \leq T_p \leq +\frac{P}{2}$ in steps of 1 day. There were no acceptable solutions for $P_3$, which can therefore be excluded. There were 9132 acceptable solutions for $P_1$ compared to 173 for $P_2$; hence $P_1$ is the preferred period but $P_2$ cannot as of now be completely excluded. A sample of 60 acceptable solutions for $P_1$ and 20 for $P_2$ are plotted against the data in Figure \ref{fig:orbits_alpha2}. 

In Table \ref{table:orbital_fits} we report the median and the uncertainties (defined as the 2.3\% and 97.7\% percentiles) of the acceptable solutions for both $P_1$ and $P_2$. Note that due to the lack of radial velocity data, the degenerate solutions with $\Omega \rightarrow \Omega + 180^{\degr}$ and $\omega \rightarrow 180^{\degr}$ are also allowed. 

As expected, some orbital parameters are better constrained than others. Interestingly, for both possible periods $i$ and $\Omega$ are quite well-constrained and consistent with each other, so that the orientation of the orbital plane appears to be rather tightly constrained by the data despite the limited orbital coverage.  

\subsection{Isochrone masses}
\label{subsec:isochrone}

From the VLTI/GRAVITY acquisition camera image (Figure \ref{fig:alpha_cru_main}, IV) we calculated the H band flux ratio between $\alpha$ Cru B and A through aperture photometry, finding 

\begin{align}
\left. \frac{f_{B}}{f_A} \right|_{\mathrm{H \: band}} = 75.4 \pm 2.8 \%
\end{align}

From the 2MASS H band magnitude of the system $H_{A+B}=1.328\pm0.282$ \citep{Skrutskie2006}, we then have 

\begin{align}
H_A = 1.94\pm0.30 \\
H_B = 2.24\pm0.30 
\end{align}

\noindent Using the VLTI/PIONIER interferometric H band flux ratios

\begin{align}
\left. \frac{f_{Ab}}{f_{Aa}} \right|_{\mathrm{H \: band}} = 12.7 \pm 0.1 \% \\
\left. \frac{f_{Bb}}{f_{Ba}} \right|_{\mathrm{H \: band}} = 56.3 \pm 0.1 \%
\end{align}

\noindent -- where we used the result for the best epoch (2018-02-04) for Aa+Ab -- we can then calculate the individual H band magnitudes of each component 

\begin{align}
H_{Aa} = 2.07 \pm 0.31 \\
H_{Ab} = 4.31 \pm 0.31 \\
H_{Ba} = 2.72 \pm 0.30 \\
H_{Bb} = 3.35 \pm 0.30 
\end{align}

Even though we have a precise distance $d=106.7\pm1.5 \text{ pc}$, we are still limited by the relatively large error in the 2MASS magnitude of this very bright star. 

We can overcome this by appealing to stellar evolution tracks. Specifically, we make use of \texttt{MIST} isochrones \citep{Dotter16,Choi16,Paxton11,Paxton13,Paxton15} with solar metallicity and initial rotational velocity of 0.4 times the critical velocity. Further constraints come from the Tycho2 magnitudes \citep{Hog2000} of A and B 

\begin{align}
V_{T,A} = 1.245 \pm 0.013 \\
V_{T,B} = 1.547 \pm 0.016 \\ 
B_{T,A} = 1.053 \pm 0.016 \\
B_{T,B} = 1.366 \pm 0.014 
\end{align}

\noindent, the spectroscopic temperature of Aa $T_{Aa} \approx 29 \text{ kK}$ and the fact that all components should be co-eval. We neglect interstellar extinction since $\alpha$ Cru is relatively nearby and still approximately within the local bubble (e.g. the 3D dust map in \cite{Wang2025} gives $A_V \sim 0.02 \pm 0.01$). 

\textit{Alpha Cru A}: For a given H band magnitude of the system $H_A$ within 1.34 and 2.54 ($\pm 2\sigma$), we computed the individual H band magnitudes from the interferometric flux ratio. For each isochrone, we find the corresponding stars and choose the isochrone for which the primary temperature is the closest to the spectroscopic tempature $T_{Aa} \approx 29 \text{ kK}$. We then compare the corresponding combined $V_T$ and $B_T$ magnitudes with the measured ones from Tycho2. We found $H_A=2.13$ for a best fit isochrone of 7 Myr. The corresponding photometric masses, radii and temperatures are reported in Table \ref{table:isochrone_results}. The photometric masses agree with the dynamical masses (Table \ref{table:orbital_fits}) within their uncertainties. The isochrone age is somewhat younger than the estimated age of the Acrux cluster (11 Myr) but a more careful investigation of the systematic uncertainties of both estimates is needed to assess whether this discrepancy is significant. We note that fixing the isochrone age to 11 Myr does not work because \texttt{MIST} isochrones cannot yield a temperature higher than 26,000 K for such an age. 

\begin{table}
\centering
\caption{\label{table:isochrone_results} Isochrone fitting results to Alpha Crucis photometry.}
\begin{tabular}{ccccc}
\hline \hline

& \shortstack{Masses\\($M_{\odot}$)} & \shortstack{Radii\\($R_{\odot}$)} & \shortstack{Temperature\\($K$)} & \shortstack{Age\\(Myr)} \\ [0.3cm]

$\alpha$ Cru Aa+Ab & \shortstack{15.3\\6.5} & \shortstack{6.8\\3.2} & \shortstack{28,950$^*$\\19,750} & 7.1 \\ [0.3cm]

$\alpha$ Cru Ba+Bb & \shortstack{12.4\\9.8} & \shortstack{5.4\\4.4} & \shortstack{26,950\\24,350} & - \\ [0.3cm]

$\alpha$ Cru Ca+Cb & \shortstack{4.5\\$\gtrsim 0.64$} & \shortstack{2.5\\-} & \shortstack{15,900\\-} & - \\ [0.3cm]

\hline
\end{tabular}
\tablenotetext{0}{* Fixed to match spectroscopic temperature.}
\end{table}

\textit{Alpha Cru B}: Based on the the analysis of Alpha Cru A we have that $H_B = 2.43$ ($H_A$ and $H_B$ are perfectly correlated). The VLTI/PIONIER interferometric flux ratio yields the individual H band magnitudes. Requiring the isochrone to be the same as Alpha Cru A, we can then find the photometric masses $M_{Ba}\simeq12.4 M_{\odot}$ and $M_{Bb}\simeq9.8 M_{\odot}$. The photometric radii and temperatures are also reported in Table \ref{table:isochrone_results}. Spectral observations of $\alpha$ Cru B are encouraged in order to test our photometric temperatures. 

\textit{Alpha Cru C}: Using the best fit isochrone for Alpha Cru A we can also estimate the masses for Alpha Cru Ca+Cb. Assuming the flux contribution of Cb is small (\cite{Hernandez79} did not find any sign of it in their spectra) and using the Tycho2 $V_T=4.799\pm0.009$ and $B_T=4.644\pm0.014$ magnitudes of Alpha Cru C, we find $M_{Ca}\simeq 4.5 M_{\odot}$. The mass function of the SB1 solution in \cite{Hernandez79} 

\begin{align}
\frac{M_{Cb}^3}{(M_{Cb}+M_{Ca})^2} \sin^3 i = 0.010 M_{\odot}
\end{align}

\noindent then implies a minimum mass $M_{Cb} \gtrsim 0.64 M_{\odot}$ for the companion. The photometric temperature and radius of Ca are also reported in Table \ref{table:isochrone_results}. 

\section{Discussion}
\label{sec:discussion}

\subsection{Orbital misalignment}

Despite the orbit of Ba+Bb being still not fully constrained, the orientation of its orbital plane is already rather tightly constrained. This allows to calculate the mutual inclination betwen the orbits Aa+Ab and Ba+Bb

\begin{align}
\begin{split}
&i_{\mathrm{mut}} = \\&\arccos (\cos i_{\mathrm{A}} \cos i_{\mathrm{B}} + \sin i_\mathrm{A} \sin i_{\mathrm{B}} \cos (\Omega_\mathrm{A} - \Omega_{\mathrm{B}}) )
\end{split}
\end{align}

\noindent which is $i_{\mathrm{mut}} = 48.9 \pm 4.5^{\degr}$ or $136.1_{-4.7}^{+7.2}$ if $P_B = 405.4 \text{ days}$ and $i_{\mathrm{mut}} = 49.5 \pm 4.0^{\degr}$ or $138.0_{-4.2}^{+4.9}$ if $P_B = 202.9 \text{ days}$ (there are two possibilities for each period due to the $(\Omega,\omega) \leftrightarrow (\Omega + 180^{\degr}, \omega + 180^{\degr})$ degeneracy of the orbital solutions due to the lack of radial velocities). Therefore, we can conclude that the two orbits are rather significantly misaligned. It would be interesting to also measure the mutual inclination between the inner binaries and the outer orbit A+B, but given the period $P_{A+B} \simeq 1.3 \text{ kyr}$ this will require a very long monitoring time. 

The misalignment is a hint for a dynamical formation scenario for the inner quadruple of Alpha Crucis. The system was likely formed by dynamical unfolding in an unstable multiple system. 

\subsection{Very high multiplicity}

The discovery of component Bb upgrades the multiplicity of Alpha Crucis to a septuple. Figure \ref{fig:schematic} summarizes our current knowledge of the entire Alpha Crucis system. The total mass in the main quadruple A+B is $M_{A+B} \simeq 46.3 M_{\odot}$ and the total mass in the system is $M \simeq 52 M_{\odot}$. 

It is interesting to note that the multiplicity of $\alpha$ Cru is significantly higher than the bias-corrected companion frequency of early B-type stars ($\mathrm{CF}=1.8\pm0.4$) and even of O-type stars ($\mathrm{CF}=2.1\pm0.3$) estimated from surveys combining high-angular resolution and spectroscopy observations \citep[][and references therein]{Offner2023}. 

At the typical distances $d \gtrsim 1$ kpc for the massive stars in such surveys $\alpha$ Cru A+B would appear as an unresolved quadruple system for spectroscopic observations and as a binary or triple system to interferometric observations depending on the resolution and field of view (Aa+Ab would be unresolved). It is questionable whether its quadruple nature could be unveiled even with dedicated observations and analysis. It is also questionable whether the faint M0V companion to $\alpha$ Cru C could be unveiled were it at a projected separation of just 0.2". This raises the possibility that the multiplicity of massive stars is still underesimated. A dedicated literature search and follow up observations of the massive stars in the Sco-Cen association is probably warranted to test this possibility. 

\section{Conclusions and future work}
\label{sec:conclusion}

In this paper we have done a deeper investigation of the quadruple system at the heart of the first-magnitude star Alpha Crucis, which happens to be the brightest and closest very high multiplicity massive star to the Sun. Our results can be summarized as follows:

\begin{enumerate}
\item A combination of interferometric and radial velocity data allowed for a complete orbital solution for the previously known spectroscopic binary Aa+Ab. In particular, we find dynamical masses $M_{Aa} = 17.2 \pm 1.2 M_{\odot}$ and $M_{Ab} = 6.8 \pm 0.3 M_{\odot}$. The photometric masses of Aa and Ab are consistent with these values. 

\item We presented a spectral atlas for Alpha Cru A with more than 250 lines identified. The spectrum constrains the temperature $T_{Aa} \simeq 29 \text{ kK}$ and projected rotatinal velocity $v \sin i \simeq 84 \text{ km}\text{ s}^{-1}$ of the primary star in the system. 

\item The interferometric data showed that Alpha Cru B is also a close binary, which updates the multiplicity of the whole system to a septuple. While the data is still insufficient to tightly constrain all the orbital parameters, we found that the orbital period of Ba+Bb is most likely $P\simeq 405.4 \text{ days}$, although the second harmonic $P\simeq 203 \text{ days}$ cannot yet be fully excluded. The photometric masses are $M_{Ba} \simeq 12.4 M_{\odot}$ and $M_{Bb} \simeq 9.8 M_{\odot}$. The total mass of the Alpha Crucis system is about $52 M_{\odot}$. 

\item The orbits Aa+Ab and Ba+Bb are significantly misaligned by either about $50^{\degr}$ or $137^{\degr}$, which points to a dynamical formation scenario for the quadruple. 
\end{enumerate}

Further interferometric observations of Ba+Bb should easily allow for a full orbital solution including a confirmation of the orbital period. The discrepancy between the age of our best fit \texttt{MIST} isochorne (7 Myr) and the previously inferred age of the Acrux cluster (11 Myr) should also be investigated. Finally, the very high multiplicity of Alpha Crucis compared to the bias-corrected companion frequency of massive stars inferred from surveys of more distant stars should also be considered. 

\section*{Acknowledgments}
This research has made use of CDS Astronomical Databases SIMBAD and VIZIER, NASA's Astrophysics Data System Bibliographic Services, NumPy \citep{van2011numpy} and matplotlib, a Python library for publication quality graphics \citep{Hunter2007}. 

\section*{Data availability}
The data underlying this article is publicly available from the ESO archive. Reduced data can be provided by the authors upon request. 

\bibliographystyle{aasjournal}
\bibliography{main}{}

\appendix

\section{A. VLTI observations and best-fit parameters}
\label{app:vlti_observations}

Tables \ref{table:observations_Alpha1} and \ref{table:observations_Alpha2} show details of the VLTI observations of Alpha Crucis A and B as well as the best-fit binary model parameters for each epoch. 

\begin{table*}
\centering
\caption{\label{table:observations_Alpha1} VLTI observations of Alpha Crucis A and best-fit binary model parameters for A$a$+A$b$.}
\begin{tabular}{cccccccccc}
\hline \hline
\shortstack{date\\MJD} & Instrument & \shortstack{seeing\\@500 nm\\(")} & $\frac{\lambda}{\Delta \lambda}$ & \shortstack{AT configuration} & \shortstack{$B_{\mathrm{proj,max}}$ (m) \\$\theta_{\mathrm{max}}$ (mas)} & \shortstack{calibrator\\$\theta_{\mathrm{diam}}$ (mas)} & \shortstack{$\frac{f_{\mathrm{A_b}}}{f_{\mathrm{A_a}}}$ (\%)} & \shortstack{$\Delta \alpha_*$\\(mas)} & \shortstack{$\Delta \delta$\\(mas)} \\ [0.3cm]

\shortstack{2018-02-03\\58152.39} & \shortstack{PIONIER\\H band} & 0.9 & 40 & A0-B2-J2-J3 & \shortstack{128.7\\2.46} & \shortstack{HD116244\\K0III\\1.5} & $15.4\pm0.1$ & $-6.76\pm0.05$ & $-0.65\pm0.2$ \\ [0.3cm]

\shortstack{2018-02-04\\58153.38} & \shortstack{PIONIER\\H band} & 1.0 & 40 & A0-G1-J2-J3 & \shortstack{129.0\\2.45} & \shortstack{HD116244\\K0III\\1.5} & $12.7\pm0.1$ & $-6.23\pm0.01$ & $0.64\pm0.01$ \\ [0.3cm]

\shortstack{2018-03-03\\58180.34} & \shortstack{GRAVITY\\single field\\K band} & 0.6-0.8 & 4000 & A0-G1-J2-J3 & \shortstack{126.7\\3.6} & \shortstack{HD120913\\K2III\\1.7} & $13.3\pm0.5$ & $4.59\pm0.11$ & $3.44\pm0.08$ \\ [0.3cm]

\shortstack{2019-03-27\\58569.15} & \shortstack{GRAVITY\\dual field\\K band} & 0.6 & 500 & A0-G1-J2-J3 & \shortstack{31.9\\14.2} & \shortstack{HD127193\\K1III\\0.92} & \shortstack{$13.3$\\fixed} & $-1.20\pm0.14$ & $-4.67\pm0.05$ \\ [0.3cm]

\hline
\end{tabular}
\end{table*}

\begin{table*}
\centering
\caption{\label{table:observations_Alpha2} VLTI observations of Alpha Crucis B and best-fit binary model parameters for B$a$+B$b$.}
\begin{tabular}{cccccccccc}
\hline \hline
\shortstack{date\\MJD} & Instrument & \shortstack{seeing\\@500 nm\\(")} & $\frac{\lambda}{\Delta \lambda}$ & \shortstack{AT config.} & \shortstack{$B_{\mathrm{proj,max}}$ (m) \\$\theta_{\mathrm{max}}$ (mas)} & \shortstack{calibrator\\$\theta_{\mathrm{diam}}$ (mas)} & \shortstack{$\frac{f_{\mathrm{B_b}}}{f_{\mathrm{B_a}}}$ (\%)} & \shortstack{$\Delta \alpha_*$\\(mas)} & \shortstack{$\Delta \delta$\\(mas)} \\ [0.3cm]

\shortstack{2018-02-06\\58155.37} & \shortstack{PIONIER\\H band} & 0.4 & 40 & A0-G1-J2-J3 & \shortstack{129.1\\2.45} & \shortstack{HD116244\\K0III\\1.5} & $56.3\pm0.1$ & $9.682\pm0.002$ & $-2.173\pm0.005$ \\ [0.3cm]

\shortstack{2018-03-03\\58180.39} & \shortstack{GRAVITY\\single field\\K band} & 0.7-1.1 & 4000 & A0-G1-J2-J3 & \shortstack{123.5\\3.7} & \shortstack{HD120913\\K2III\\1.7} & $56.2\pm1.2$ & $16.70\pm0.01$ & $-1.40\pm0.02$ \\ [0.3cm]

\shortstack{2018-03-04\\58181.35} & \shortstack{GRAVITY\\single field\\K band} & 0.6-0.7 & 4000 & A0-G1-J2-J3 & \shortstack{126.0\\3.6} & \shortstack{HD120913\\K2III\\1.7} & $57.2\pm0.2$ & $16.914\pm0.006$ & $-1.375\pm0.016$ \\ [0.3cm]

\shortstack{2018-03-04\\58181.40} & \shortstack{GRAVITY\\dual field\\K band} & 0.7 & 500 & A0-G1-J2-J3 & \shortstack{123.2\\3.7} & \shortstack{-} & \shortstack{56.5\\fixed} & $16.96\pm0.01$ & $-1.37\pm0.01$ \\ [0.3cm]

\shortstack{2019-03-26\\58568.18} & \shortstack{GRAVITY\\dual field\\K band} & 0.5-0.7 & 500 & A0-B2-D0-C1 & \shortstack{31.4\\14.5} & \shortstack{-} & \shortstack{56.5\\fixed} & $11.82\pm0.04$ & $-1.82\pm0.03$ \\ [0.3cm]

\shortstack{2019-03-27\\58569.13} & \shortstack{GRAVITY\\dual field\\K band} & 0.5-0.6 & 500 & A0-B2-D0-C1 & \shortstack{32.0\\14.2} & \shortstack{HD127193\\K1III\\0.92} & $56.8\pm0.6$ & $12.14\pm0.01$ & $-1.89\pm0.03$ \\ [0.3cm]

\hline
\end{tabular}
\end{table*}

\begin{figure*}[]
\centering
\includegraphics[width=\textwidth]{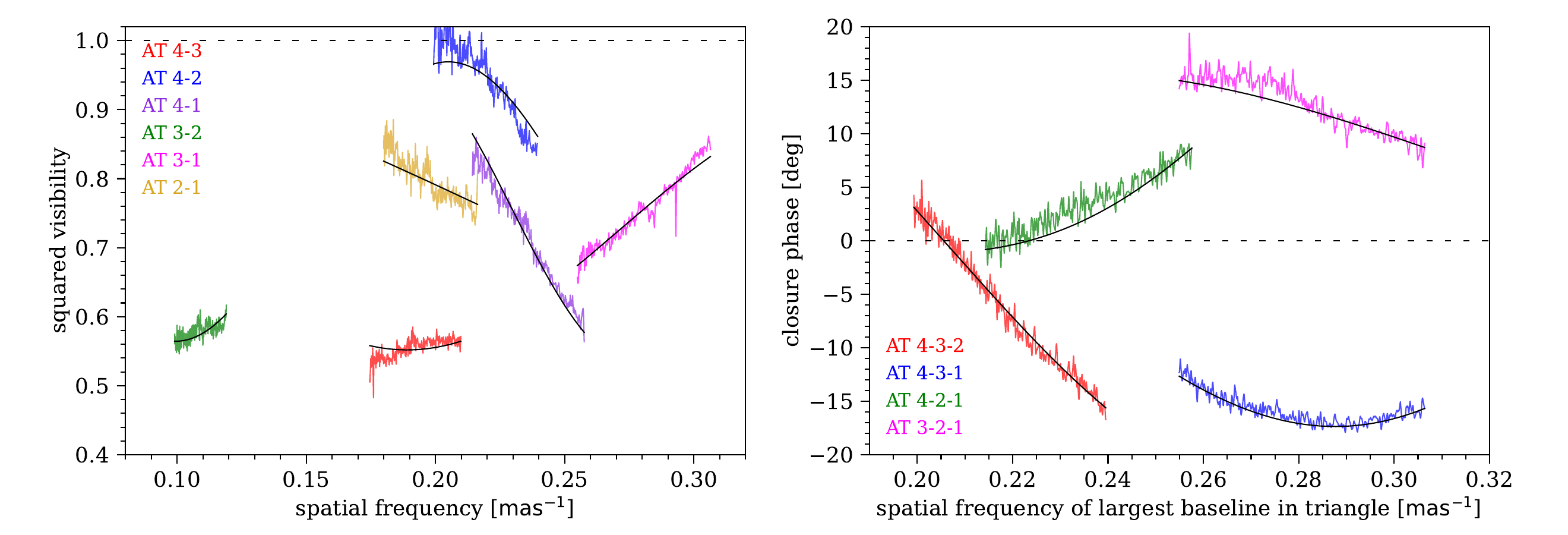}\\
\includegraphics[width=\textwidth]{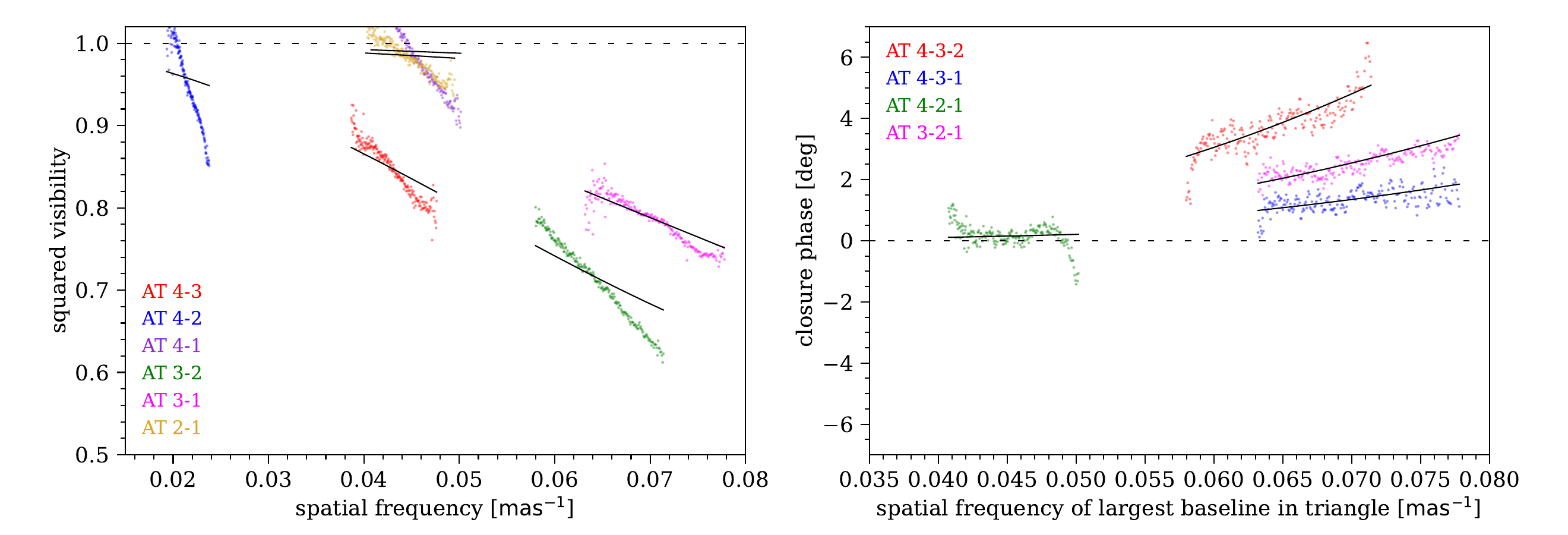}\\
\caption{\label{fig:gravity_fits_Alpha1} VLTI/GRAVITY data (colored) for Alpha Crucis Aa+Ab and best fit binary model (solid black) for Epochs 2018-03-03 (top) and 2019-03-27 (bottom). The dashed lines show the expected values for a single unresolved star.}
\end{figure*}

\begin{figure*}[]
\centering
\includegraphics[width=\textwidth]{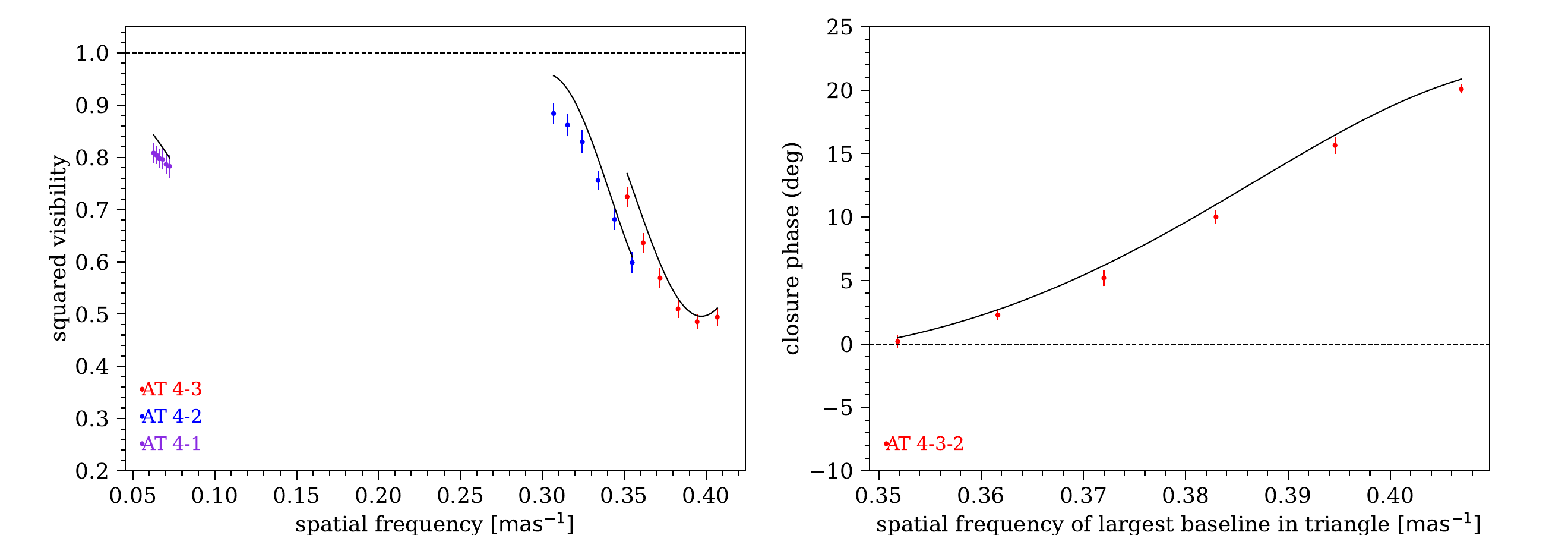}\\
\includegraphics[width=\textwidth]{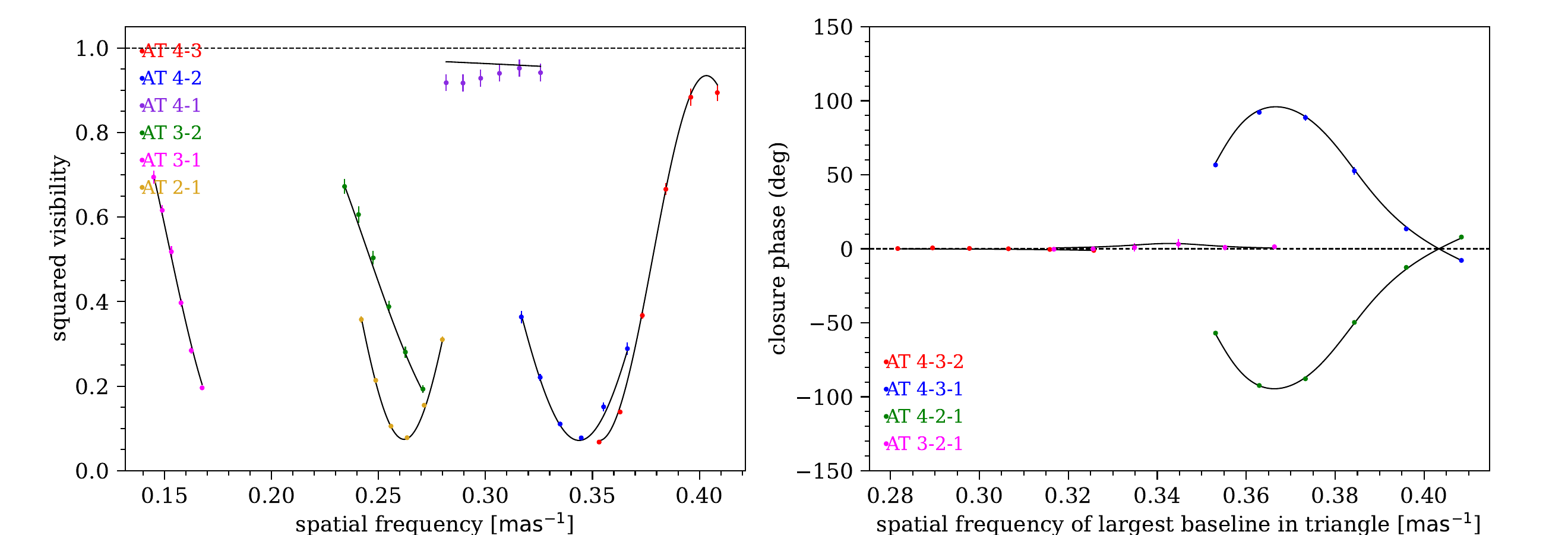}\\
\caption{\label{fig:pionier_fits_Alpha1} VLTI/PIONIER data (colored) for Alpha Crucis Aa+Ab and best fit binary model (solid black) for Epochs 2018-02-03 (top) and 2018-02-04 (bottom). The dashed lines show the expected values for a single unresolved star.}
\end{figure*}

\begin{figure*}[]
\centering
\includegraphics[width=\textwidth]{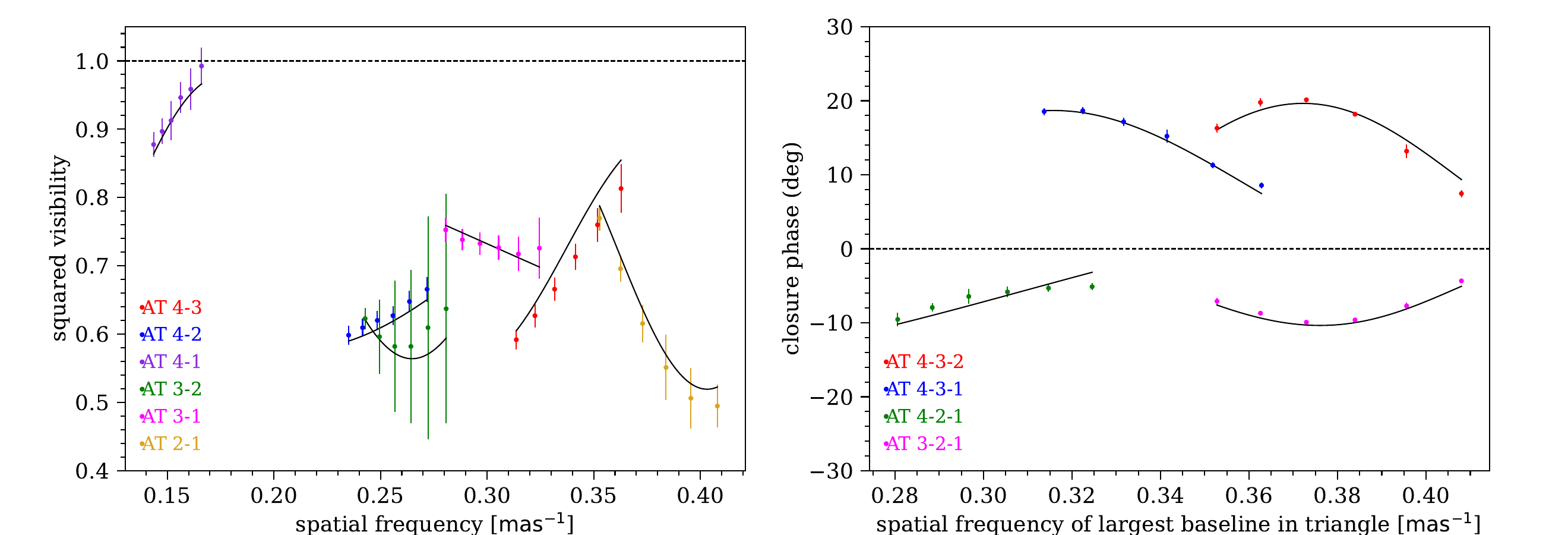}\\
\caption{\label{fig:pionier_fits_Alpha2} VLTI/PIONIER data (colored) for Alpha Crucis Ba+Bb and best fit binary model (solid black) for Epoch 2018-02-06. The dashed lines show the expected values for a single unresolved star.}
\end{figure*}

\begin{figure*}[]
\centering
\includegraphics[width=\textwidth]{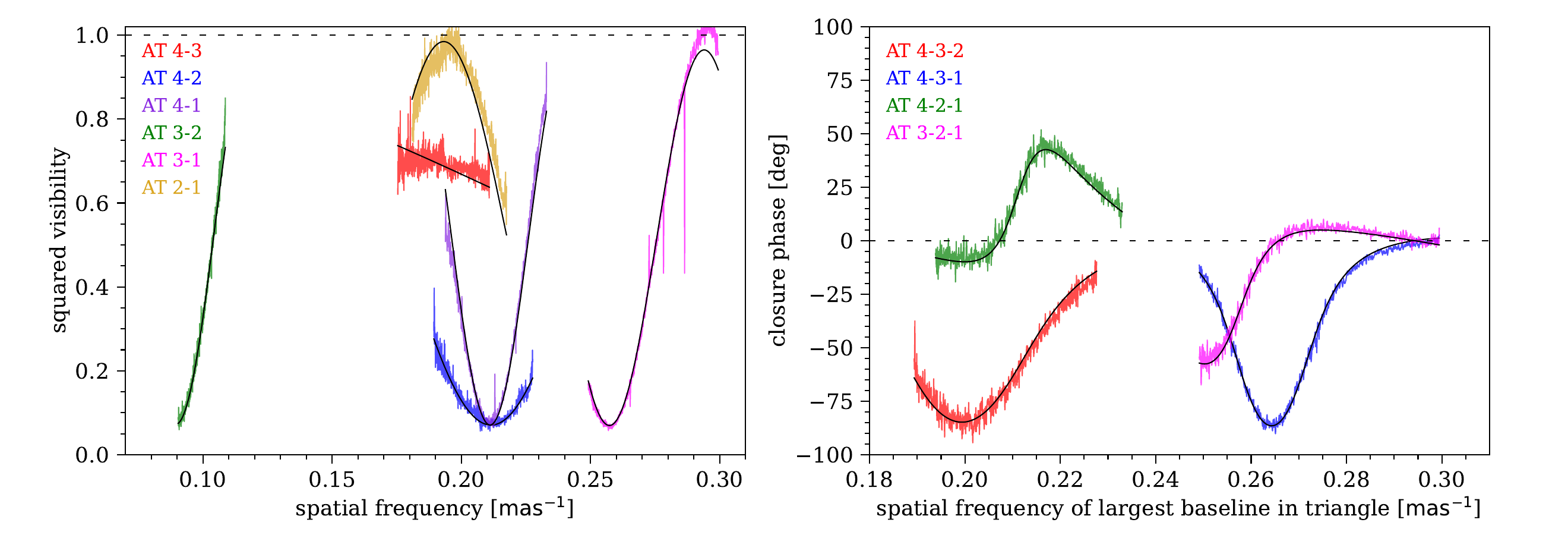}\\
\includegraphics[width=\textwidth]{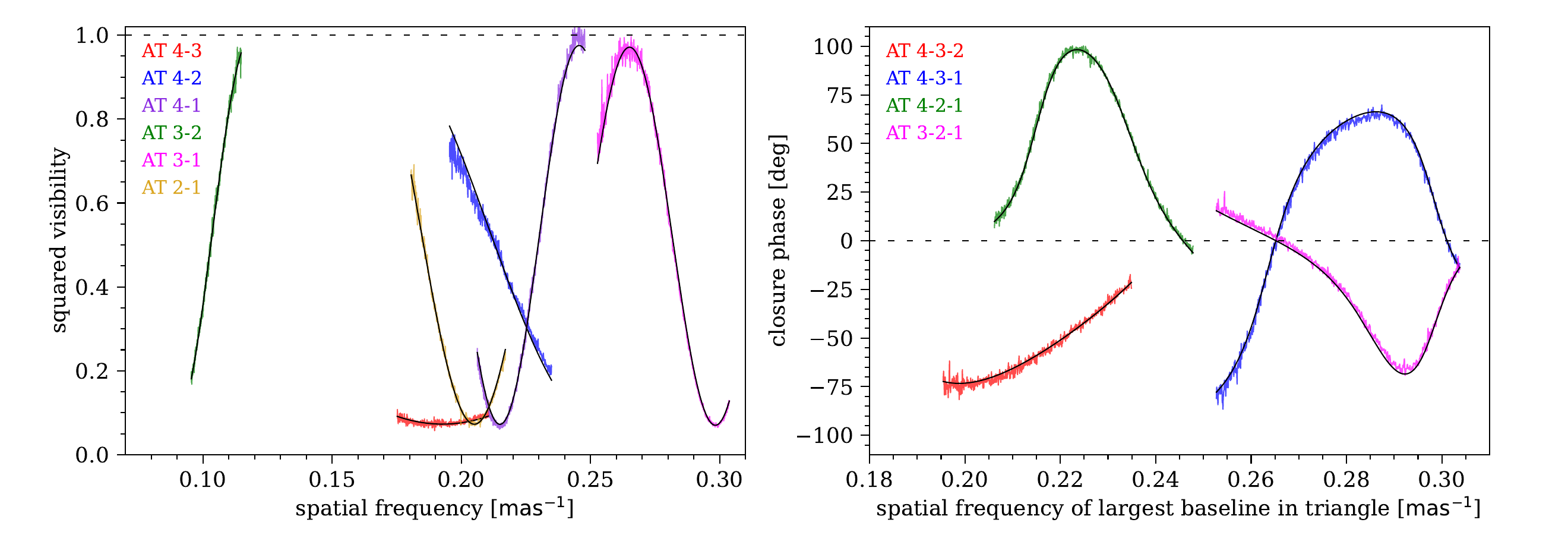}\\
\includegraphics[width=\textwidth]{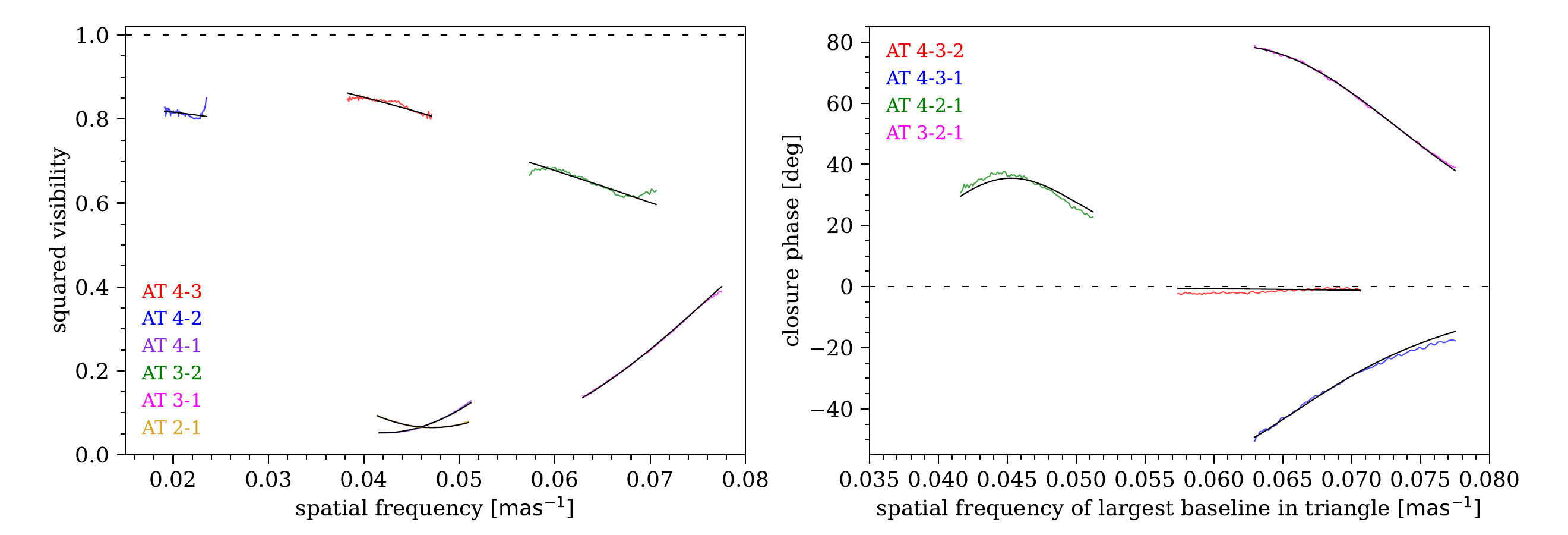} \\
\includegraphics[width=\textwidth]{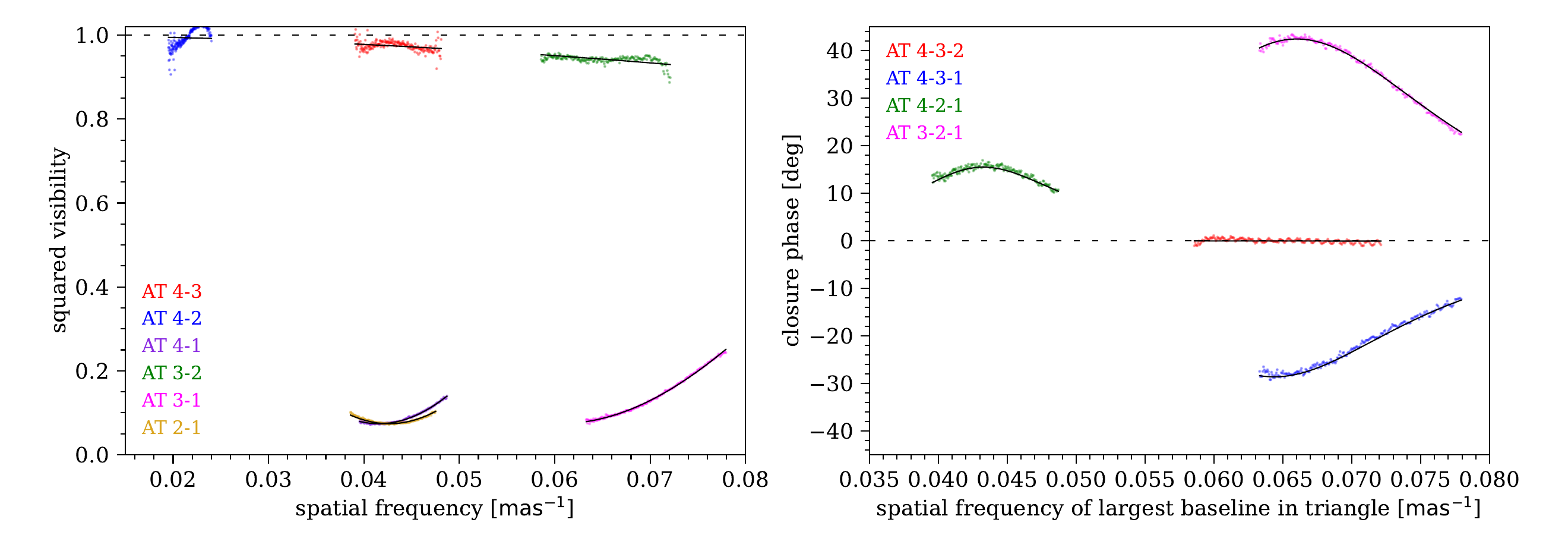}\\
\caption{\label{fig:gravity_fits_Alpha2} VLTI/GRAVITY data (colored) for Alpha Crucis Ba+Bb and best fit binary model (solid black) for Epochs 2018-03-03, 2018-03-04, 2019-03-26 and 2019-03-27 (top to bottom). The dashed lines show the expected values for a single unresolved star.}
\end{figure*}

Figures \ref{fig:gravity_fits_Alpha1} and \ref{fig:pionier_fits_Alpha1} show the interferometric data (colored) and best-fit binary model (black) for the VLTI/GRAVITY and VLTI/PIONIER observations of Alpha Cru A. Figures \ref{fig:gravity_fits_Alpha2} and \ref{fig:pionier_fits_Alpha2} are the corresponding figures for Alpha Cru B.

\begin{figure*}[]
\centering
\includegraphics[width=\textwidth]{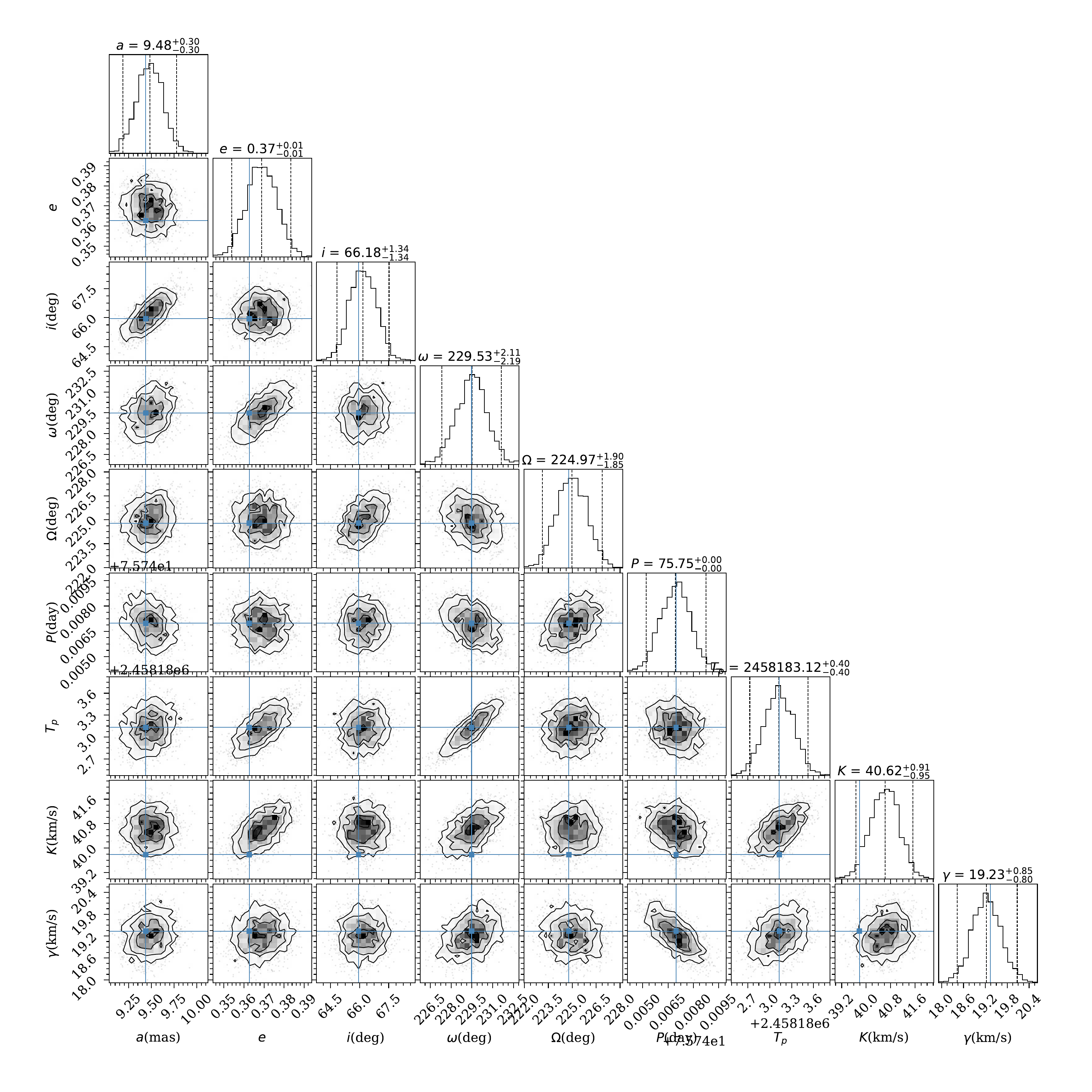}
\caption{\label{fig:corner_alpha1} Orbital parameters distributions for Alpha Crucis A.}
\end{figure*}

The full distributions of the best fit orbital parameters for $\alpha$ Cru A are shown in Figure \ref{fig:corner_alpha1}. The values marked by the blue lines show the best fit parameters for the original data, while the distributions refer to the best fit parameters of the 20,000 resamples of the data. 

\section{B. Spectra details}

Table \ref{table:spectra} summarizes the spectral observations of Alpha Cru A. 

\begin{table*}
\centering
\caption{\label{table:spectra} Details of spectral observations of Alpha Cru A.}
\begin{tabular}{cccccccc}
\hline \hline
target & \shortstack{date\\MJD} & Instrument & $\lambda$ & $\frac{\lambda}{\Delta \lambda}$ & Exposure time & \shortstack{$v_{\mathrm{helio}}$\\($\text{ km}\text{ s}^{-1}$)} & \shortstack{$v_{\mathrm{radial,Aa}}$\\($\text{ km}\text{ s}^{-1}$)} \\ [0.3cm]

$\alpha$ Cru A & \shortstack{2006-01-12\\53747.372} & UVES & \shortstack{3300-4500{\AA}\\4800-6800{\AA}} & \shortstack{65,030\\74,450} & 5 s & +17.20 & $-0.6 \pm 2.4$ \\ [0.3cm]

$\alpha$ Cru A & \shortstack{2006-04-10\\53835.164} & UVES & \shortstack{3050-3850{\AA}\\4800-6800{\AA}} & \shortstack{58,640\\87,410} & \shortstack{12.6 s\\7.5 s} & +6.90 & $13.7\pm3.0$ \\ [0.3cm]

$\alpha$ Cru A & \shortstack{2016-04-06\\57484.977} & HARPS & 3780-6900{\AA} & \shortstack{115,000} & 1 s & - & $44.3 \pm 4.5$ \\ [0.3cm]

$\alpha$ Cru A & \shortstack{2004-05-03\\53128.090} & FEROS & 3500-9200{\AA} & 48,000 & 20 s & - & $-12.1\pm2.7$ \\ [0.3cm]

\hline
\end{tabular}
\end{table*}

Table \ref{table:line_list} lists the transitions identified in the 3440-6750{\AA} spectrum of Alpha Cru A. 

\clearpage

\begin{longtable}{cccc}
\caption{\label{table:line_list} Identified spectral lines in Alpha Cru A.} \\
\toprule
Ion & Wavelength (\AA) & & Wavelength (\AA) \\
\midrule
\endfirsthead
\toprule
Ion & Wavelength (\AA) & & Wavelength (\AA) \\
\midrule
\endhead
\bottomrule
\endfoot
\midrule
\multicolumn{4}{l}{\textbf{H I}} \\
H16 & 3703.85 & H15 & 3711.97 \\
H14 & 3721.94 & H13 & 3734.37 \\
H12 & 3750.15 & H11 & 3770.63 \\
H10 & 3797.90 & H9 & 3835.38 \\
H8 & 3889.05 & H$\epsilon$ & 3970.07 \\
H$\delta$ & 4101.735 & H$\gamma$ & 4340.463 \\
H$\beta$ & 4861.325 & H$\alpha$ & 6562.80 \\
\midrule
\multicolumn{4}{l}{\textbf{HeI}} \\
HeI & 3447.59 & & 3498.64 \\
HeI & 3512.51 & & 3530.49 \\
HeI & 3554.41 & & 3587.27 \\
HeI & 3613.64 & & 3634.23 \\
HeI & 3705.00 & & 3732.88 \\
HeI & 3805.74 & & 3819.60 \\
HeI & 3833.57 & & 3867.48 \\
HeI & 3871.69 & & 3926.41 \\
HeI & 3964.73 & & 4009.26 \\
HeI & 4026.20 & & 4120.82 \\
HeI & 4143.76 & & 4168.97 \\
HeI & 4387.93 & & 4437.55 \\
HeI & 4471.50 & & 4713.17 \\
HeI & 4921.93 & & 5015.70 \\
HeI & 5047.74 & & 5875.64 \\
HeI & 6678.15 & & \\
\midrule 
\multicolumn{4}{l}{\textbf{HeII}} \\
HeII & 4199.80 & & 4541.6 \\
HeII & 4685.68 & & 5411.5 \\
\midrule
\multicolumn{4}{l}{\textbf{CII}} \\
CII & 3590.82 & & 3876.4 \\
CII & 4156.33 & & 4267.26 \\
CII & 4289.88 & & 4372.35 \\
CII & 4372.49 & & 4618.40 \\
CII & 4619.23 & & 4620.35 \\
CII & 5145.16 & & 6578.01 \\
CII & 6582.88 & &  \\
\midrule
\multicolumn{4}{l}{\textbf{NII}} \\
NII & 3995.00 & & 4035.08 \\
NII & 4041.31 & & 4236.91 \\
NII & 4241.78 & & 4379.59 \\
NII & 4530.41 & & 4630.54 \\
NII & 4447.03 & & 4601.48 \\
NII & 4613.87 & & 4621.39 \\
NII & 4643.09 & & 4654.53 \\
NII & 4803.29 & & 5001.13 \\
NII & 5001.48 & & 5002.70 \\
NII & 5005.15 & & 5007.33 \\
NII & 5010.62 & & 5011.31 \\
NII & 5012.04 & & 5045.10 \\
NII & 5679.56 & & 5686.21 \\
NII & 5710.77 & &  \\
\midrule
\multicolumn{4}{l}{\textbf{OII}} \\
OII & 3470.28 & & 3470.68 \\
OII & 3727.32 & & 3759.06 \\
OII & 3761.40 & & 3762.47 \\
OII & 3777.42 & & 3847.89 \\
OII & 3850.80 & & 3851.03 \\
OII & 3851.47 & & 3856.62 \\
OII & 3882.45 & & 3911.97 \\
OII & 3919.27 & & 3945.04 \\
OII & 3954.36 & & 3973.26 \\
OII & 3982.72 & & 4060.60 \\
OII & 4061.03 & & 4069.62 \\
OII & 4069.88 & & 4072.15 \\
OII & 4075.86 & & 4078.84 \\
OII & 4085.11 & & 4092.93 \\
OII & 4094.14 & & 4095.64 \\
OII & 4096.19 & & 4096.53 \\
OII & 4096.93 & & 4097.26 \\
OII & 4098.24 & & 4110.79 \\
OII & 4112.02 & & 4129.32 \\
OII & 4132.80 & & 4153.23 \\
OII & 4156.53 & & 4185.44 \\
OII & 4189.79 & & 4276 \\
OII & 4291.25 & & 4294.78 \\
OII & 4303.61 & & 4317.14 \\
OII & 4319.63 & & 4325.77 \\
OII & 4345.56 & & 4349.43 \\
OII & 4366.89 & & 4371.62 \\
OII & 4378.03 & & 4378.43 \\
OII & 4395.94 & & 4397.86 \\
OII & 4414.90 & & 4416.97 \\
OII & 4443.00 & & 4443.52 \\
OII & 4452.38 & & 4465.42 \\
OII & 4466.24 & & 4466.42 \\
OII & 4467.48 & & 4467.72 \\
OII & 4467.92 & & 4469.37 \\
OII & 4488.20 & & 4491.23 \\
OII & 4590.97 & & 4596.18 \\
OII & 4609.44 & & 4610.20 \\
OII & 4638.86 & & 4641.81 \\
OII & 4649.14 & & 4650.84 \\
OII & 4661.63 & & 4673.73 \\
OII & 4676.24 & & 4699.00 \\
OII & 4699.22 & & 4705.35 \\
OII & 4710.01 & & 4843.37 \\
OII & 4844.91 & & 4890.86 \\
OII & 4906.88 & & 4924.53 \\
OII & 4941.07 & & 4943.00 \\
OII & 4955.71 & & 5159.94 \\
OII & 5206.65 & & 6641.05 \\
OII & 6721.36 & &  \\
\midrule
\multicolumn{4}{l}{\textbf{NeII}} \\
NeII & 3480.72 & & 3481.93 \\
NeII & 3542.24 & & 3542.85 \\
NeII & 3543.79 & & 3568.50 \\
NeII & 3572.02 & & 3572.38 \\
NeII & 3574.61 & & 3601.06 \\
NeII & 3602.78 & & 3626.54 \\
NeII & 3628.03 & & 3643.93 \\
NeII & 3644.86 & & 3651.26 \\
NeII & 3652.81 & & 3664.07 \\
NeII & 3694.21 & & 3756.39 \\
NeII & 3760.51 & &  \\
\midrule
\multicolumn{4}{l}{\textbf{MgII}} \\
MgII & 4481.21 & &  \\
\midrule
\multicolumn{4}{l}{\textbf{AlIII}} \\
AlIII & 5696.60 & & 5722.73 \\
\midrule
\multicolumn{4}{l}{\textbf{SiIII}} \\
SiIII & 3486.91 & & 3791.41 \\
SiIII & 4219.96 & & 4552.62 \\
SiIII & 4567.82 & & 4574.76 \\
SiIII & 4813.33 & & 4819.72 \\
SiIII & 4828.95 & & 5739.73 \\
\midrule
\multicolumn{4}{l}{\textbf{SiIV}} \\
SiIV & 4088.85 & & 4116.10 \\
SiIV & 4212.41 & & 4654.32 \\
\midrule
\multicolumn{4}{l}{\textbf{SIII}} \\
SIII & 3497.29 & & 3709.33 \\
SIII & 3710.41 & & 4253.50 \\
SIII & 4284.89 & & 4332.65 \\
SIII & 4354.52 & & 4361.47 \\
SIII & 5160.08 & & 5219.32 \\
\midrule
\multicolumn{4}{l}{\textbf{CaII}} \\
i.s. CaII & 3933.66 & &  \\
\end{longtable}

\end{document}